\PassOptionsToPackage{dvipsnames,table}{xcolor}
\documentclass[conference]{IEEEtran}
\usepackage{fancyhdr}

\usepackage{amsmath,amssymb,amsfonts}
\usepackage{graphicx}
\usepackage{textcomp}
\usepackage[numbers]{natbib}

\makeatletter
\newcommand{\linebreakand}{%
      \end{@IEEEauthorhalign}
      \hfill\mbox{}\par
      \mbox{}\hfill\begin{@IEEEauthorhalign}
}
\makeatother

\newcommand{\Description}[1]{}
\def\BibTeX{{\rm B\kern-.05em{\sc i\kern-.025em b}\kern-.08em
T\kern-.1667em\lower.7ex\hbox{E}\kern-.125emX}}

\usepackage{xspace}
\usepackage{graphicx}
\usepackage[normalem]{ulem}%
\usepackage{amsmath}
\usepackage[inline]{enumitem}
\setlist[enumerate]{label=(\arabic*)}
\usepackage{natbib}
\setlist{nosep}%

\usepackage{hyperref}
\hypersetup{linkcolor=black,citecolor=black,anchorcolor=black,filecolor=black,menucolor=black,runcolor=black,urlcolor=black,hidelinks}
\usepackage{breakurl}

\usepackage{array}
\usepackage{booktabs}
\usepackage{multirow}
\usepackage{makecell}
\usepackage{ragged2e}
\usepackage{tabularx}
\usepackage{pifont}

\usepackage[font=small, aboveskip=2.0pt, belowskip=2.0pt]{caption}%
\usepackage{subcaption}
\usepackage{fancybox}
\usepackage{wrapfig}
\makeatletter
\def\input@path{{./dependencies/}{./}}
\makeatother
\usepackage{tikz-uml}
\usepackage{pgfplots}
\pgfplotsset{compat=1.18}

\pgfplotsset{
      EvalPlotStyleOpenMP/.style={orange!90!black, thick, dashed, mark=x, mark options={scale=0.78}},
      EvalPlotStylePkdbOpenMP/.style={green!60!black, thick, densely dotted, mark=*, mark options={scale=0.72}},
      EvalPlotStyleCuda/.style={purple!80!black, very thick, solid, mark=o, mark options={scale=0.72}},
      EvalPlotStylePkdbCuda/.style={blue!80!black, thick, dotted, mark=square, mark options={solid, scale=0.65}},
      EvalPlotStyleHIP/.style={red!75!black, thick, dashed, mark=diamond, mark options={solid, scale=0.86}},
      EvalPlotStylePkdbHIP/.style={teal!70!black, thick, dashdotted, mark=triangle, mark options={solid, scale=0.86}},
      EvalPlotStyleBaseline/.style={black, thick, densely dashed, mark=triangle, mark options={solid, scale=0.72}},
      EvalBarOpenMP/.style={fill=orange!80!black},
      EvalBarPkdbOpenMP/.style={fill=green!60!black},
      EvalBarCuda/.style={fill=purple!80!black},
      EvalBarPkdbCuda/.style={fill=blue!70!black},
      EvalBarStartup/.style={fill=gray!50},
      EvalBarLineTracing/.style={fill=red!70!black},
      EvalBarDispatch/.style={fill=orange!80!black},
      EvalBarPyKokkosExec/.style={fill=yellow!80!black},
      EvalBarKernelExec/.style={fill=blue!70!black},
      EvalBarAppExec/.style={fill=green!60!black},
}
\pgfplotsset{
      /pgfplots/legend image code/.code={
                  \draw[mark repeat=2,mark phase=2,#1]
                  plot coordinates {
                              (0cm,0cm)
                              (0.21cm,0cm)
                              (0.44cm,0cm)
                        };
            },
}

\usepackage{listings}

\usepackage{tikz}
\usetikzlibrary{arrows.meta, positioning, calc}

\usepackage{mathpartir}

\usepackage{adjustbox}

\usepackage{algorithm}
\usepackage{algpseudocode}
\usepackage{longtable}

\usepackage{flushend}

\newcommand{\XSpace}[1]{}
\newcommand{\XComment}[1]{}
\newcommand{\CodeIn}[1]{\texttt{\small{#1}}}
\newcommand{\MyPara}[1]{\par\vskip 2pt\noindent\textbf{#1}.}
\newcommand{\DefMacro}[2]{\expandafter\newcommand\csname rmk-#1\endcsname{#2}}
\newcommand{\UseMacro}[1]{\csname rmk-#1\endcsname}
\newcommand{\UseMacroRound}[2]{%
      \begingroup
      \edef\rmk@val{\UseMacro{#1}}%
      \pgfmathprintnumber[fixed, precision=#2, zerofill]{\rmk@val}%
      \endgroup
}

\newcommand{\ourTool}{\textsc{pkdb}\xspace}

\newcommand{\Numba}{Numba\xspace}
\newcommand{\Triton}{Triton\xspace}
\newcommand{\PyCuda}{PyCuda\xspace}
\newcommand{\CUDA}{CUDA\xspace}
\newcommand{\Cuda}{\CUDA}
\newcommand{\OpenMP}{OpenMP\xspace}
\newcommand{\Openmp}{\OpenMP}
\newcommand{\HIP}{HIP\xspace}
\newcommand{\Hip}{\HIP}
\newcommand{\Python}{Python\xspace}
\newcommand{\CUDAGDB}{CUDA-GDB\xspace}
\newcommand{\ROCGDB}{ROCgdb\xspace}
\newcommand{\GDB}{GDB\xspace}
\newcommand{\PDB}{pdb\xspace}
\newcommand{\pdbp}{$\text{\PDB}^\star$\xspace}
\newcommand{\SlowdownRatio}{$t^{\ourTool}/t^{\text{pdb}}$\xspace}
\newcommand{\NVIDIA}{NVIDIA\xspace}

\newcommand{\AMD}{AMD\xspace}
\newcommand{\Intel}{Intel\xspace}

\newcommand{\PyKokkos}{PyKokkos\xspace}
\newcommand{\Kokkos}{Kokkos\xspace}
\newcommand{\CPU}{CPU\xspace}
\newcommand{\CPUs}{CPUs\xspace}
\newcommand{\GPU}{GPU\xspace}
\newcommand{\GPUs}{GPUs\xspace}

\newcommand{\HPC}{HPC\xspace}
\newcommand{\pybind}{pybind11\xspace}
\newcommand{\cupy}{CuPy\xspace}

\newcommand{\eDSL}{eDSL\xspace}
\newcommand{\NumPy}{NumPy\xspace}
\newcommand{\CuPy}{CuPy\xspace}
\newcommand{\Vista}{V7\xspace}

\newcommand{\Tokyo}{Local\xspace}
\newcommand{\amdCluster}{AMC\xspace}
\newcommand{\ourToolGitGub}{\url{https://github.com/EngineeringSoftware/pkdb}}
\newcommand{\RTXFiveKAda}{NVIDIA RTX~5000 Ada\xspace}
\newcommand{\parallelOpsCode}{\CodeIn{parallel\_\textless op\textgreater}\xspace}

\newcommand{\mbeddr}{mbeddr C\xspace}
\newcommand{\Legion}{Legion\xspace}

\newcommand{\liveeval}{live code evaluation\xspace}
\newcommand{\Liveeval}{Live code evaluation\xspace}
\newcommand{\kernelSubstitute}{kernel call site substitution\xspace}
\newcommand{\KernelSubstitute}{Kernel call site substitution\xspace}
\newcommand{\EditDebug}{edit-and-debug\xspace}

\newcommand{\Exa}{ExaMiniMD\xspace}
\newcommand{\Boltz}{Boltzmann\xspace}
\newcommand{\Ewald}{Ewald\xspace}

\newcommand{\EwaldIndex}{528\xspace}

\newcommand{\chartDashedLineOpacity}{45}

\newcommand{\substFromFunc}{f}
\newcommand{\substToFunc}{t}

\newcommand{\substituteTo}{$\substToFunc$\xspace}
\newcommand{\substituteFrom}{$\substFromFunc$\xspace}
\newcommand{\substGlobalMap}{$\mathcal{M}_{\text{glob}}$\xspace}
\newcommand{\substLocalMap}{$\mathcal{M}_{\text{loc}}$\xspace}
\newcommand{\substNewState}{$\mathcal{H}$\xspace}
\newcommand{\newStateGetS}{$\mathcal{H}[\substFromFunc]$\xspace}
\newcommand{\substMapping}{$\substFromFunc \rightarrow \substToFunc$\xspace}
\newcommand{\substLineFunc}{\ell}
\newcommand{\substLocalMapEntry}{$\mathcal{M}_{\text{loc}}[\substFromFunc,\substLineFunc]$\xspace}
\newcommand{\substLine}{$\substLineFunc$\xspace}
\newcommand{\substResultFunc}{r}
\newcommand{\substResult}{$\substResultFunc$\xspace}
\newcommand{\argMismatch}{"Signature mismatch error"\xspace}
\newcommand{\CodeLine}{\textit{code}}

\IfFileExists{tables/eval-hotswap-macros.tex}{
\DefMacro{eval-hotswap-temperature-name}{Temperature}
\DefMacro{eval-hotswap-temperature-tA}{87.11}
\DefMacro{eval-hotswap-temperature-tA1}{43.76}
\DefMacro{eval-hotswap-temperature-tA2}{43.34}
\DefMacro{eval-hotswap-temperature-tB}{66.55}
\DefMacro{eval-hotswap-temperature-speedup}{1.31}
\DefMacro{eval-hotswap-temperature-hotswap-cmd}{8.63}

\DefMacro{eval-hotswap-kine-name}{KinE}
\DefMacro{eval-hotswap-kine-tA}{88.25}
\DefMacro{eval-hotswap-kine-tA1}{44.41}
\DefMacro{eval-hotswap-kine-tA2}{43.84}
\DefMacro{eval-hotswap-kine-tB}{67.18}
\DefMacro{eval-hotswap-kine-speedup}{1.31}
\DefMacro{eval-hotswap-kine-hotswap-cmd}{0.65}

\DefMacro{eval-hotswap-nve-initial-name}{NVE initial}
\DefMacro{eval-hotswap-nve-initial-tA}{89.64}
\DefMacro{eval-hotswap-nve-initial-tA1}{44.90}
\DefMacro{eval-hotswap-nve-initial-tA2}{44.75}
\DefMacro{eval-hotswap-nve-initial-tB}{67.78}
\DefMacro{eval-hotswap-nve-initial-speedup}{1.32}
\DefMacro{eval-hotswap-nve-initial-hotswap-cmd}{6.21}

\DefMacro{eval-hotswap-nve-final-name}{NVE final}
\DefMacro{eval-hotswap-nve-final-tA}{89.08}
\DefMacro{eval-hotswap-nve-final-tA1}{44.22}
\DefMacro{eval-hotswap-nve-final-tA2}{44.86}
\DefMacro{eval-hotswap-nve-final-tB}{67.77}
\DefMacro{eval-hotswap-nve-final-speedup}{1.31}
\DefMacro{eval-hotswap-nve-final-hotswap-cmd}{6.27}

\DefMacro{eval-hotswap-total-name}{Mean}
\DefMacro{eval-hotswap-total-tA}{88.52}
\DefMacro{eval-hotswap-total-tA1}{44.32}
\DefMacro{eval-hotswap-total-tA2}{44.20}
\DefMacro{eval-hotswap-total-tB}{67.32}
\DefMacro{eval-hotswap-total-speedup}{1.31}
\DefMacro{eval-hotswap-total-hotswap-cmd}{5.44}

\DefMacro{eval-hotswap-meta-system-suffix}{x86\_64\_nvidia\_rtx\_5000\_ada\_generation}
}{}
\IfFileExists{tables/eval-benchmark-macros.tex}{
\DefMacro{eval-bench-examinimd-debugwall-pkdbopenmp-t-min}{0.6046}
\DefMacro{eval-bench-examinimd-debugwall-pkdbopenmp-t-max}{1.2272}
\DefMacro{eval-bench-examinimd-debugwall-pkdbcuda-t-min}{1.2088}
\DefMacro{eval-bench-examinimd-debugwall-pkdbcuda-t-max}{1.3582}
\DefMacro{eval-bench-examinimd-debugwall-baseline-t-min}{11.5704}
\DefMacro{eval-bench-examinimd-debugwall-baseline-t-max}{460.8648}
\DefMacro{eval-bench-examinimd-debugwall-amd8xmi300x-pkdbhip-t-y-1}{1.1022}
\DefMacro{eval-bench-examinimd-debugwall-amd8xmi300x-pkdbhip-t-y-2}{1.1291}
\DefMacro{eval-bench-examinimd-debugwall-amd8xmi300x-pkdbhip-t-y-3}{1.131}
\DefMacro{eval-bench-examinimd-debugwall-amd8xmi300x-pkdbhip-t-y-4}{1.1109}
\DefMacro{eval-bench-examinimd-debugwall-amd8xmi300x-pkdbhip-t-y-5}{1.1351}
\DefMacro{eval-bench-examinimd-debugwall-amd8xmi300x-baseline-t-y-1}{13.5286}
\DefMacro{eval-bench-examinimd-debugwall-amd8xmi300x-baseline-t-y-2}{27.5406}
\DefMacro{eval-bench-examinimd-debugwall-amd8xmi300x-baseline-t-y-3}{56.4738}
\DefMacro{eval-bench-examinimd-debugwall-amd8xmi300x-baseline-t-y-4}{88.2528}
\DefMacro{eval-bench-examinimd-debugwall-amd8xmi300x-baseline-t-y-5}{505.4981}

\DefMacro{eval-max-slowdown-openmp}{2.183128}
\DefMacro{eval-max-slowdown-cuda}{2.334481}
\DefMacro{eval-max-slowdown-hip}{2.140189}

\DefMacro{eval-bench-examinimd-exec-openmp-max-avg-pkdb-slowdown}{1.745378}
\DefMacro{eval-bench-examinimd-exec-cuda-max-avg-pkdb-slowdown}{1.948299}
\DefMacro{eval-bench-examinimd-exec-hip-max-avg-pkdb-slowdown}{1.535331}
\DefMacro{eval-bench-boltzmann-exec-openmp-max-avg-pkdb-slowdown}{2.183128}
\DefMacro{eval-bench-boltzmann-exec-cuda-max-avg-pkdb-slowdown}{2.334481}
\DefMacro{eval-bench-boltzmann-exec-hip-max-avg-pkdb-slowdown}{2.140189}
\DefMacro{eval-bench-ewald-exec-openmp-max-avg-pkdb-slowdown}{1.703424}
\DefMacro{eval-bench-ewald-exec-cuda-max-avg-pkdb-slowdown}{1.858831}
\DefMacro{eval-bench-ewald-exec-hip-max-avg-pkdb-slowdown}{1.843157}

\DefMacro{eval-bench-boltzmann-exec-amd-mi300x-hip-pdb-y-1}{8.1177}
\DefMacro{eval-bench-boltzmann-exec-amd-mi300x-hip-pdb-y-2}{10.8257}
\DefMacro{eval-bench-boltzmann-exec-amd-mi300x-hip-pdb-y-3}{22.7876}
\DefMacro{eval-bench-boltzmann-exec-amd-mi300x-hip-pdb-y-4}{38.4516}
\DefMacro{eval-bench-boltzmann-exec-amd-mi300x-hip-pdb-y-5}{164.4288}
\DefMacro{eval-bench-boltzmann-exec-amd-mi300x-hip-pdb-y-6}{306.2601}
\DefMacro{eval-bench-boltzmann-exec-amd-mi300x-hip-pdb-min}{8.1177}
\DefMacro{eval-bench-boltzmann-exec-amd-mi300x-hip-pdb-max}{306.2601}
\DefMacro{eval-bench-boltzmann-exec-amd-mi300x-hip-pkdb-y-1}{10.8455}
\DefMacro{eval-bench-boltzmann-exec-amd-mi300x-hip-pkdb-y-2}{18.0961}
\DefMacro{eval-bench-boltzmann-exec-amd-mi300x-hip-pkdb-y-3}{50.8378}
\DefMacro{eval-bench-boltzmann-exec-amd-mi300x-hip-pkdb-y-4}{90.9005}
\DefMacro{eval-bench-boltzmann-exec-amd-mi300x-hip-pkdb-y-5}{417.4305}
\DefMacro{eval-bench-boltzmann-exec-amd-mi300x-hip-pkdb-y-6}{826.8663}
\DefMacro{eval-bench-boltzmann-exec-amd-mi300x-hip-pkdb-min}{10.8455}
\DefMacro{eval-bench-boltzmann-exec-amd-mi300x-hip-pkdb-max}{826.8663}
\DefMacro{eval-bench-boltzmann-exec-amd-mi300x-hip-avg-pkdb-slowdown}{2.140189}
\DefMacro{eval-bench-boltzmann-exec-amd-mi300x-hip-min-pkdb-slowdown}{1.336031}
\DefMacro{eval-bench-boltzmann-exec-amd-mi300x-hip-max-pkdb-slowdown}{2.699883}
\DefMacro{eval-bench-boltzmann-exec-amd-mi300x-hip-median-pkdb-slowdown}{2.297483}

\DefMacro{eval-bench-boltzmann-exec-vista-gh200-cuda-pdb-y-1}{3.1593}
\DefMacro{eval-bench-boltzmann-exec-vista-gh200-cuda-pdb-y-2}{5.5321}
\DefMacro{eval-bench-boltzmann-exec-vista-gh200-cuda-pdb-y-3}{14.0223}
\DefMacro{eval-bench-boltzmann-exec-vista-gh200-cuda-pdb-y-4}{30.8947}
\DefMacro{eval-bench-boltzmann-exec-vista-gh200-cuda-pdb-y-5}{143.2983}
\DefMacro{eval-bench-boltzmann-exec-vista-gh200-cuda-pdb-y-6}{285.0903}
\DefMacro{eval-bench-boltzmann-exec-vista-gh200-cuda-pdb-min}{3.1593}
\DefMacro{eval-bench-boltzmann-exec-vista-gh200-cuda-pdb-max}{285.0903}
\DefMacro{eval-bench-boltzmann-exec-vista-gh200-cuda-pkdb-y-1}{8.1409}
\DefMacro{eval-bench-boltzmann-exec-vista-gh200-cuda-pkdb-y-2}{12.6528}
\DefMacro{eval-bench-boltzmann-exec-vista-gh200-cuda-pkdb-y-3}{29.8607}
\DefMacro{eval-bench-boltzmann-exec-vista-gh200-cuda-pkdb-y-4}{64.1577}
\DefMacro{eval-bench-boltzmann-exec-vista-gh200-cuda-pkdb-y-5}{292.3325}
\DefMacro{eval-bench-boltzmann-exec-vista-gh200-cuda-pkdb-y-6}{576.1444}
\DefMacro{eval-bench-boltzmann-exec-vista-gh200-cuda-pkdb-min}{8.1409}
\DefMacro{eval-bench-boltzmann-exec-vista-gh200-cuda-pkdb-max}{576.1444}
\DefMacro{eval-bench-boltzmann-exec-vista-gh200-cuda-avg-pkdb-slowdown}{2.188514}
\DefMacro{eval-bench-boltzmann-exec-vista-gh200-cuda-min-pkdb-slowdown}{2.020919}
\DefMacro{eval-bench-boltzmann-exec-vista-gh200-cuda-max-pkdb-slowdown}{2.576805}
\DefMacro{eval-bench-boltzmann-exec-vista-gh200-cuda-median-pkdb-slowdown}{2.103086}
\DefMacro{eval-bench-boltzmann-exec-vista-gh200-openmp-pdb-y-1}{3.5603}
\DefMacro{eval-bench-boltzmann-exec-vista-gh200-openmp-pdb-y-2}{6.9963}
\DefMacro{eval-bench-boltzmann-exec-vista-gh200-openmp-pdb-y-3}{19.1192}
\DefMacro{eval-bench-boltzmann-exec-vista-gh200-openmp-pdb-y-4}{42.613}
\DefMacro{eval-bench-boltzmann-exec-vista-gh200-openmp-pdb-y-5}{201.5397}
\DefMacro{eval-bench-boltzmann-exec-vista-gh200-openmp-pdb-y-6}{399.706}
\DefMacro{eval-bench-boltzmann-exec-vista-gh200-openmp-pdb-min}{3.5603}
\DefMacro{eval-bench-boltzmann-exec-vista-gh200-openmp-pdb-max}{399.706}
\DefMacro{eval-bench-boltzmann-exec-vista-gh200-openmp-pkdb-y-1}{9.428}
\DefMacro{eval-bench-boltzmann-exec-vista-gh200-openmp-pkdb-y-2}{16.2801}
\DefMacro{eval-bench-boltzmann-exec-vista-gh200-openmp-pkdb-y-3}{39.6921}
\DefMacro{eval-bench-boltzmann-exec-vista-gh200-openmp-pkdb-y-4}{87.0894}
\DefMacro{eval-bench-boltzmann-exec-vista-gh200-openmp-pkdb-y-5}{403.5235}
\DefMacro{eval-bench-boltzmann-exec-vista-gh200-openmp-pkdb-y-6}{800.1118}
\DefMacro{eval-bench-boltzmann-exec-vista-gh200-openmp-pkdb-min}{9.428}
\DefMacro{eval-bench-boltzmann-exec-vista-gh200-openmp-pkdb-max}{800.1118}
\DefMacro{eval-bench-boltzmann-exec-vista-gh200-openmp-avg-pkdb-slowdown}{2.183128}
\DefMacro{eval-bench-boltzmann-exec-vista-gh200-openmp-min-pkdb-slowdown}{2.001751}
\DefMacro{eval-bench-boltzmann-exec-vista-gh200-openmp-max-pkdb-slowdown}{2.648091}
\DefMacro{eval-bench-boltzmann-exec-vista-gh200-openmp-median-pkdb-slowdown}{2.059881}

\DefMacro{eval-bench-boltzmann-exec-tokyo-cuda-pdb-y-1}{2.9878}
\DefMacro{eval-bench-boltzmann-exec-tokyo-cuda-pdb-y-2}{5.5846}
\DefMacro{eval-bench-boltzmann-exec-tokyo-cuda-pdb-y-3}{14.288}
\DefMacro{eval-bench-boltzmann-exec-tokyo-cuda-pdb-y-4}{31.3384}
\DefMacro{eval-bench-boltzmann-exec-tokyo-cuda-pdb-y-5}{145.1235}
\DefMacro{eval-bench-boltzmann-exec-tokyo-cuda-pdb-y-6}{283.3477}
\DefMacro{eval-bench-boltzmann-exec-tokyo-cuda-pdb-min}{2.9878}
\DefMacro{eval-bench-boltzmann-exec-tokyo-cuda-pdb-max}{283.3477}
\DefMacro{eval-bench-boltzmann-exec-tokyo-cuda-pkdb-y-1}{8.8019}
\DefMacro{eval-bench-boltzmann-exec-tokyo-cuda-pkdb-y-2}{14.1358}
\DefMacro{eval-bench-boltzmann-exec-tokyo-cuda-pkdb-y-3}{31.9734}
\DefMacro{eval-bench-boltzmann-exec-tokyo-cuda-pkdb-y-4}{67.0034}
\DefMacro{eval-bench-boltzmann-exec-tokyo-cuda-pkdb-y-5}{301.8399}
\DefMacro{eval-bench-boltzmann-exec-tokyo-cuda-pkdb-y-6}{587.665}
\DefMacro{eval-bench-boltzmann-exec-tokyo-cuda-pkdb-min}{8.8019}
\DefMacro{eval-bench-boltzmann-exec-tokyo-cuda-pkdb-max}{587.665}
\DefMacro{eval-bench-boltzmann-exec-tokyo-cuda-avg-pkdb-slowdown}{2.334481}
\DefMacro{eval-bench-boltzmann-exec-tokyo-cuda-min-pkdb-slowdown}{2.074007}
\DefMacro{eval-bench-boltzmann-exec-tokyo-cuda-max-pkdb-slowdown}{2.945947}
\DefMacro{eval-bench-boltzmann-exec-tokyo-cuda-median-pkdb-slowdown}{2.18792}
\DefMacro{eval-bench-boltzmann-exec-tokyo-openmp-pdb-y-1}{6.035}
\DefMacro{eval-bench-boltzmann-exec-tokyo-openmp-pdb-y-2}{10.5241}
\DefMacro{eval-bench-boltzmann-exec-tokyo-openmp-pdb-y-3}{32.9884}
\DefMacro{eval-bench-boltzmann-exec-tokyo-openmp-pdb-y-4}{77.2272}
\DefMacro{eval-bench-boltzmann-exec-tokyo-openmp-pdb-y-5}{350.2575}
\DefMacro{eval-bench-boltzmann-exec-tokyo-openmp-pdb-y-6}{684.1961}
\DefMacro{eval-bench-boltzmann-exec-tokyo-openmp-pdb-min}{6.035}
\DefMacro{eval-bench-boltzmann-exec-tokyo-openmp-pdb-max}{684.1961}
\DefMacro{eval-bench-boltzmann-exec-tokyo-openmp-pkdb-y-1}{10.6977}
\DefMacro{eval-bench-boltzmann-exec-tokyo-openmp-pkdb-y-2}{20.6842}
\DefMacro{eval-bench-boltzmann-exec-tokyo-openmp-pkdb-y-3}{53.7291}
\DefMacro{eval-bench-boltzmann-exec-tokyo-openmp-pkdb-y-4}{119.968}
\DefMacro{eval-bench-boltzmann-exec-tokyo-openmp-pkdb-y-5}{568.9305}
\DefMacro{eval-bench-boltzmann-exec-tokyo-openmp-pkdb-y-6}{1127.9736}
\DefMacro{eval-bench-boltzmann-exec-tokyo-openmp-pkdb-min}{10.6977}
\DefMacro{eval-bench-boltzmann-exec-tokyo-openmp-pkdb-max}{1127.9736}
\DefMacro{eval-bench-boltzmann-exec-tokyo-openmp-avg-pkdb-slowdown}{1.698854}
\DefMacro{eval-bench-boltzmann-exec-tokyo-openmp-min-pkdb-slowdown}{1.553442}
\DefMacro{eval-bench-boltzmann-exec-tokyo-openmp-max-pkdb-slowdown}{1.965413}
\DefMacro{eval-bench-boltzmann-exec-tokyo-openmp-median-pkdb-slowdown}{1.638669}

\DefMacro{eval-bench-ewald-exec-amd-mi300x-hip-pdb-y-1}{6.0319}
\DefMacro{eval-bench-ewald-exec-amd-mi300x-hip-pdb-y-2}{6.873}
\DefMacro{eval-bench-ewald-exec-amd-mi300x-hip-pdb-y-3}{9.4004}
\DefMacro{eval-bench-ewald-exec-amd-mi300x-hip-pdb-y-4}{15.9178}
\DefMacro{eval-bench-ewald-exec-amd-mi300x-hip-pdb-y-5}{17.0787}
\DefMacro{eval-bench-ewald-exec-amd-mi300x-hip-pdb-y-6}{18.1807}
\DefMacro{eval-bench-ewald-exec-amd-mi300x-hip-pdb-y-7}{20.5303}
\DefMacro{eval-bench-ewald-exec-amd-mi300x-hip-pdb-y-8}{21.786}
\DefMacro{eval-bench-ewald-exec-amd-mi300x-hip-pdb-y-9}{22.7156}
\DefMacro{eval-bench-ewald-exec-amd-mi300x-hip-pdb-y-10}{24.0061}
\DefMacro{eval-bench-ewald-exec-amd-mi300x-hip-pdb-y-11}{24.8978}
\DefMacro{eval-bench-ewald-exec-amd-mi300x-hip-pdb-y-12}{25.4753}
\DefMacro{eval-bench-ewald-exec-amd-mi300x-hip-pdb-y-13}{47.0118}
\DefMacro{eval-bench-ewald-exec-amd-mi300x-hip-pdb-y-14}{91.1682}
\DefMacro{eval-bench-ewald-exec-amd-mi300x-hip-pdb-y-15}{219.7932}
\DefMacro{eval-bench-ewald-exec-amd-mi300x-hip-pdb-min}{6.0319}
\DefMacro{eval-bench-ewald-exec-amd-mi300x-hip-pdb-max}{219.7932}
\DefMacro{eval-bench-ewald-exec-amd-mi300x-hip-pkdb-y-1}{20.4757}
\DefMacro{eval-bench-ewald-exec-amd-mi300x-hip-pkdb-y-2}{20.8559}
\DefMacro{eval-bench-ewald-exec-amd-mi300x-hip-pkdb-y-3}{23.8308}
\DefMacro{eval-bench-ewald-exec-amd-mi300x-hip-pkdb-y-4}{30.0776}
\DefMacro{eval-bench-ewald-exec-amd-mi300x-hip-pkdb-y-5}{31.2786}
\DefMacro{eval-bench-ewald-exec-amd-mi300x-hip-pkdb-y-6}{32.5738}
\DefMacro{eval-bench-ewald-exec-amd-mi300x-hip-pkdb-y-7}{34.6706}
\DefMacro{eval-bench-ewald-exec-amd-mi300x-hip-pkdb-y-8}{35.9907}
\DefMacro{eval-bench-ewald-exec-amd-mi300x-hip-pkdb-y-9}{36.8142}
\DefMacro{eval-bench-ewald-exec-amd-mi300x-hip-pkdb-y-10}{38.0332}
\DefMacro{eval-bench-ewald-exec-amd-mi300x-hip-pkdb-y-11}{38.8879}
\DefMacro{eval-bench-ewald-exec-amd-mi300x-hip-pkdb-y-12}{39.4934}
\DefMacro{eval-bench-ewald-exec-amd-mi300x-hip-pkdb-y-13}{61.7546}
\DefMacro{eval-bench-ewald-exec-amd-mi300x-hip-pkdb-y-14}{104.8094}
\DefMacro{eval-bench-ewald-exec-amd-mi300x-hip-pkdb-y-15}{230.6623}
\DefMacro{eval-bench-ewald-exec-amd-mi300x-hip-pkdb-min}{20.4757}
\DefMacro{eval-bench-ewald-exec-amd-mi300x-hip-pkdb-max}{230.6623}
\DefMacro{eval-bench-ewald-exec-amd-mi300x-hip-avg-pkdb-slowdown}{1.843157}
\DefMacro{eval-bench-ewald-exec-amd-mi300x-hip-min-pkdb-slowdown}{1.049451}
\DefMacro{eval-bench-ewald-exec-amd-mi300x-hip-max-pkdb-slowdown}{3.394569}
\DefMacro{eval-bench-ewald-exec-amd-mi300x-hip-median-pkdb-slowdown}{1.65201}

\DefMacro{eval-bench-ewald-exec-vista-gh200-cuda-pdb-y-1}{4.4651}
\DefMacro{eval-bench-ewald-exec-vista-gh200-cuda-pdb-y-2}{6.0704}
\DefMacro{eval-bench-ewald-exec-vista-gh200-cuda-pdb-y-3}{12.4076}
\DefMacro{eval-bench-ewald-exec-vista-gh200-cuda-pdb-y-4}{25.7078}
\DefMacro{eval-bench-ewald-exec-vista-gh200-cuda-pdb-y-5}{28.1102}
\DefMacro{eval-bench-ewald-exec-vista-gh200-cuda-pdb-y-6}{30.9182}
\DefMacro{eval-bench-ewald-exec-vista-gh200-cuda-pdb-y-7}{35.2593}
\DefMacro{eval-bench-ewald-exec-vista-gh200-cuda-pdb-y-8}{37.3922}
\DefMacro{eval-bench-ewald-exec-vista-gh200-cuda-pdb-y-9}{38.9823}
\DefMacro{eval-bench-ewald-exec-vista-gh200-cuda-pdb-y-10}{41.2415}
\DefMacro{eval-bench-ewald-exec-vista-gh200-cuda-pdb-y-11}{42.9198}
\DefMacro{eval-bench-ewald-exec-vista-gh200-cuda-pdb-y-12}{46.0061}
\DefMacro{eval-bench-ewald-exec-vista-gh200-cuda-pdb-y-13}{88.6552}
\DefMacro{eval-bench-ewald-exec-vista-gh200-cuda-pdb-y-14}{177.4325}
\DefMacro{eval-bench-ewald-exec-vista-gh200-cuda-pdb-y-15}{438.525}
\DefMacro{eval-bench-ewald-exec-vista-gh200-cuda-pdb-min}{4.4651}
\DefMacro{eval-bench-ewald-exec-vista-gh200-cuda-pdb-max}{438.525}
\DefMacro{eval-bench-ewald-exec-vista-gh200-cuda-pkdb-y-1}{13.0792}
\DefMacro{eval-bench-ewald-exec-vista-gh200-cuda-pkdb-y-2}{13.3938}
\DefMacro{eval-bench-ewald-exec-vista-gh200-cuda-pkdb-y-3}{20.1035}
\DefMacro{eval-bench-ewald-exec-vista-gh200-cuda-pkdb-y-4}{32.9317}
\DefMacro{eval-bench-ewald-exec-vista-gh200-cuda-pkdb-y-5}{35.5343}
\DefMacro{eval-bench-ewald-exec-vista-gh200-cuda-pkdb-y-6}{38.7979}
\DefMacro{eval-bench-ewald-exec-vista-gh200-cuda-pkdb-y-7}{42.7054}
\DefMacro{eval-bench-ewald-exec-vista-gh200-cuda-pkdb-y-8}{44.8922}
\DefMacro{eval-bench-ewald-exec-vista-gh200-cuda-pkdb-y-9}{46.3868}
\DefMacro{eval-bench-ewald-exec-vista-gh200-cuda-pkdb-y-10}{49.1444}
\DefMacro{eval-bench-ewald-exec-vista-gh200-cuda-pkdb-y-11}{51.0726}
\DefMacro{eval-bench-ewald-exec-vista-gh200-cuda-pkdb-y-12}{52.704}
\DefMacro{eval-bench-ewald-exec-vista-gh200-cuda-pkdb-y-13}{96.1108}
\DefMacro{eval-bench-ewald-exec-vista-gh200-cuda-pkdb-y-14}{183.893}
\DefMacro{eval-bench-ewald-exec-vista-gh200-cuda-pkdb-y-15}{452.0659}
\DefMacro{eval-bench-ewald-exec-vista-gh200-cuda-pkdb-min}{13.0792}
\DefMacro{eval-bench-ewald-exec-vista-gh200-cuda-pkdb-max}{452.0659}
\DefMacro{eval-bench-ewald-exec-vista-gh200-cuda-avg-pkdb-slowdown}{1.389073}
\DefMacro{eval-bench-ewald-exec-vista-gh200-cuda-min-pkdb-slowdown}{1.030878}
\DefMacro{eval-bench-ewald-exec-vista-gh200-cuda-max-pkdb-slowdown}{2.929207}
\DefMacro{eval-bench-ewald-exec-vista-gh200-cuda-median-pkdb-slowdown}{1.200577}
\DefMacro{eval-bench-ewald-exec-vista-gh200-openmp-pdb-y-1}{1.4356}
\DefMacro{eval-bench-ewald-exec-vista-gh200-openmp-pdb-y-2}{2.9401}
\DefMacro{eval-bench-ewald-exec-vista-gh200-openmp-pdb-y-3}{8.3666}
\DefMacro{eval-bench-ewald-exec-vista-gh200-openmp-pdb-y-4}{17.9333}
\DefMacro{eval-bench-ewald-exec-vista-gh200-openmp-pdb-y-5}{19.6548}
\DefMacro{eval-bench-ewald-exec-vista-gh200-openmp-pdb-y-6}{21.9798}
\DefMacro{eval-bench-ewald-exec-vista-gh200-openmp-pdb-y-7}{25.1303}
\DefMacro{eval-bench-ewald-exec-vista-gh200-openmp-pdb-y-8}{26.1508}
\DefMacro{eval-bench-ewald-exec-vista-gh200-openmp-pdb-y-9}{27.3146}
\DefMacro{eval-bench-ewald-exec-vista-gh200-openmp-pdb-y-10}{29.4499}
\DefMacro{eval-bench-ewald-exec-vista-gh200-openmp-pdb-y-11}{30.1853}
\DefMacro{eval-bench-ewald-exec-vista-gh200-openmp-pdb-y-12}{32.4591}
\DefMacro{eval-bench-ewald-exec-vista-gh200-openmp-pdb-y-13}{63.4934}
\DefMacro{eval-bench-ewald-exec-vista-gh200-openmp-pdb-y-14}{127.8842}
\DefMacro{eval-bench-ewald-exec-vista-gh200-openmp-pdb-y-15}{316.0203}
\DefMacro{eval-bench-ewald-exec-vista-gh200-openmp-pdb-min}{1.4356}
\DefMacro{eval-bench-ewald-exec-vista-gh200-openmp-pdb-max}{316.0203}
\DefMacro{eval-bench-ewald-exec-vista-gh200-openmp-pkdb-y-1}{8.2059}
\DefMacro{eval-bench-ewald-exec-vista-gh200-openmp-pkdb-y-2}{9.6605}
\DefMacro{eval-bench-ewald-exec-vista-gh200-openmp-pkdb-y-3}{15.1933}
\DefMacro{eval-bench-ewald-exec-vista-gh200-openmp-pkdb-y-4}{24.925}
\DefMacro{eval-bench-ewald-exec-vista-gh200-openmp-pkdb-y-5}{26.1871}
\DefMacro{eval-bench-ewald-exec-vista-gh200-openmp-pkdb-y-6}{28.64}
\DefMacro{eval-bench-ewald-exec-vista-gh200-openmp-pkdb-y-7}{31.9449}
\DefMacro{eval-bench-ewald-exec-vista-gh200-openmp-pkdb-y-8}{32.6792}
\DefMacro{eval-bench-ewald-exec-vista-gh200-openmp-pkdb-y-9}{34.3139}
\DefMacro{eval-bench-ewald-exec-vista-gh200-openmp-pkdb-y-10}{35.8344}
\DefMacro{eval-bench-ewald-exec-vista-gh200-openmp-pkdb-y-11}{39.6431}
\DefMacro{eval-bench-ewald-exec-vista-gh200-openmp-pkdb-y-12}{39.3883}
\DefMacro{eval-bench-ewald-exec-vista-gh200-openmp-pkdb-y-13}{70.9358}
\DefMacro{eval-bench-ewald-exec-vista-gh200-openmp-pkdb-y-14}{131.3879}
\DefMacro{eval-bench-ewald-exec-vista-gh200-openmp-pkdb-y-15}{329.6476}
\DefMacro{eval-bench-ewald-exec-vista-gh200-openmp-pkdb-min}{8.2059}
\DefMacro{eval-bench-ewald-exec-vista-gh200-openmp-pkdb-max}{329.6476}
\DefMacro{eval-bench-ewald-exec-vista-gh200-openmp-avg-pkdb-slowdown}{1.703424}
\DefMacro{eval-bench-ewald-exec-vista-gh200-openmp-min-pkdb-slowdown}{1.027397}
\DefMacro{eval-bench-ewald-exec-vista-gh200-openmp-max-pkdb-slowdown}{5.716007}
\DefMacro{eval-bench-ewald-exec-vista-gh200-openmp-median-pkdb-slowdown}{1.271171}

\DefMacro{eval-bench-ewald-exec-tokyo-cuda-pdb-y-1}{2.749}
\DefMacro{eval-bench-ewald-exec-tokyo-cuda-pdb-y-2}{3.5511}
\DefMacro{eval-bench-ewald-exec-tokyo-cuda-pdb-y-3}{7.7807}
\DefMacro{eval-bench-ewald-exec-tokyo-cuda-pdb-y-4}{15.4833}
\DefMacro{eval-bench-ewald-exec-tokyo-cuda-pdb-y-5}{16.6603}
\DefMacro{eval-bench-ewald-exec-tokyo-cuda-pdb-y-6}{17.7152}
\DefMacro{eval-bench-ewald-exec-tokyo-cuda-pdb-y-7}{20.8027}
\DefMacro{eval-bench-ewald-exec-tokyo-cuda-pdb-y-8}{22.5428}
\DefMacro{eval-bench-ewald-exec-tokyo-cuda-pdb-y-9}{23.4145}
\DefMacro{eval-bench-ewald-exec-tokyo-cuda-pdb-y-10}{24.4798}
\DefMacro{eval-bench-ewald-exec-tokyo-cuda-pdb-y-11}{25.7738}
\DefMacro{eval-bench-ewald-exec-tokyo-cuda-pdb-y-12}{26.9188}
\DefMacro{eval-bench-ewald-exec-tokyo-cuda-pdb-y-13}{52.9636}
\DefMacro{eval-bench-ewald-exec-tokyo-cuda-pdb-y-14}{104.9358}
\DefMacro{eval-bench-ewald-exec-tokyo-cuda-pdb-min}{2.749}
\DefMacro{eval-bench-ewald-exec-tokyo-cuda-pdb-max}{104.9358}
\DefMacro{eval-bench-ewald-exec-tokyo-cuda-pkdb-y-1}{12.4538}
\DefMacro{eval-bench-ewald-exec-tokyo-cuda-pkdb-y-2}{13.2055}
\DefMacro{eval-bench-ewald-exec-tokyo-cuda-pkdb-y-3}{17.6964}
\DefMacro{eval-bench-ewald-exec-tokyo-cuda-pkdb-y-4}{24.9792}
\DefMacro{eval-bench-ewald-exec-tokyo-cuda-pkdb-y-5}{26.3256}
\DefMacro{eval-bench-ewald-exec-tokyo-cuda-pkdb-y-6}{28.4515}
\DefMacro{eval-bench-ewald-exec-tokyo-cuda-pkdb-y-7}{30.7884}
\DefMacro{eval-bench-ewald-exec-tokyo-cuda-pkdb-y-8}{31.1441}
\DefMacro{eval-bench-ewald-exec-tokyo-cuda-pkdb-y-9}{32.9575}
\DefMacro{eval-bench-ewald-exec-tokyo-cuda-pkdb-y-10}{34.2069}
\DefMacro{eval-bench-ewald-exec-tokyo-cuda-pkdb-y-11}{35.3222}
\DefMacro{eval-bench-ewald-exec-tokyo-cuda-pkdb-y-12}{36.8515}
\DefMacro{eval-bench-ewald-exec-tokyo-cuda-pkdb-y-13}{63.4562}
\DefMacro{eval-bench-ewald-exec-tokyo-cuda-pkdb-y-14}{115.0816}
\DefMacro{eval-bench-ewald-exec-tokyo-cuda-pkdb-min}{12.4538}
\DefMacro{eval-bench-ewald-exec-tokyo-cuda-pkdb-max}{115.0816}
\DefMacro{eval-bench-ewald-exec-tokyo-cuda-avg-pkdb-slowdown}{1.858831}
\DefMacro{eval-bench-ewald-exec-tokyo-cuda-min-pkdb-slowdown}{1.096686}
\DefMacro{eval-bench-ewald-exec-tokyo-cuda-max-pkdb-slowdown}{4.530302}
\DefMacro{eval-bench-ewald-exec-tokyo-cuda-median-pkdb-slowdown}{1.443794}
\DefMacro{eval-bench-ewald-exec-tokyo-openmp-pdb-y-1}{2.179}
\DefMacro{eval-bench-ewald-exec-tokyo-openmp-pdb-y-2}{7.3005}
\DefMacro{eval-bench-ewald-exec-tokyo-openmp-pdb-y-3}{27.7124}
\DefMacro{eval-bench-ewald-exec-tokyo-openmp-pdb-y-4}{64.2738}
\DefMacro{eval-bench-ewald-exec-tokyo-openmp-pdb-y-5}{70.8417}
\DefMacro{eval-bench-ewald-exec-tokyo-openmp-pdb-y-6}{79.9723}
\DefMacro{eval-bench-ewald-exec-tokyo-openmp-pdb-y-7}{90.3835}
\DefMacro{eval-bench-ewald-exec-tokyo-openmp-pdb-y-8}{97.4845}
\DefMacro{eval-bench-ewald-exec-tokyo-openmp-pdb-y-9}{99.6788}
\DefMacro{eval-bench-ewald-exec-tokyo-openmp-pdb-y-10}{109.8143}
\DefMacro{eval-bench-ewald-exec-tokyo-openmp-pdb-y-11}{111.9267}
\DefMacro{eval-bench-ewald-exec-tokyo-openmp-pdb-y-12}{122.5447}
\DefMacro{eval-bench-ewald-exec-tokyo-openmp-pdb-y-13}{243.0842}
\DefMacro{eval-bench-ewald-exec-tokyo-openmp-pdb-y-14}{478.679}
\DefMacro{eval-bench-ewald-exec-tokyo-openmp-pdb-min}{2.179}
\DefMacro{eval-bench-ewald-exec-tokyo-openmp-pdb-max}{478.679}
\DefMacro{eval-bench-ewald-exec-tokyo-openmp-pkdb-y-1}{9.9784}
\DefMacro{eval-bench-ewald-exec-tokyo-openmp-pkdb-y-2}{14.9983}
\DefMacro{eval-bench-ewald-exec-tokyo-openmp-pkdb-y-3}{35.7721}
\DefMacro{eval-bench-ewald-exec-tokyo-openmp-pkdb-y-4}{73.3569}
\DefMacro{eval-bench-ewald-exec-tokyo-openmp-pkdb-y-5}{78.9188}
\DefMacro{eval-bench-ewald-exec-tokyo-openmp-pkdb-y-6}{87.2292}
\DefMacro{eval-bench-ewald-exec-tokyo-openmp-pkdb-y-7}{97.7931}
\DefMacro{eval-bench-ewald-exec-tokyo-openmp-pkdb-y-8}{103.7294}
\DefMacro{eval-bench-ewald-exec-tokyo-openmp-pkdb-y-9}{109.3781}
\DefMacro{eval-bench-ewald-exec-tokyo-openmp-pkdb-y-10}{115.0555}
\DefMacro{eval-bench-ewald-exec-tokyo-openmp-pkdb-y-11}{119.5566}
\DefMacro{eval-bench-ewald-exec-tokyo-openmp-pkdb-y-12}{126.8626}
\DefMacro{eval-bench-ewald-exec-tokyo-openmp-pkdb-y-13}{247.9178}
\DefMacro{eval-bench-ewald-exec-tokyo-openmp-pkdb-y-14}{481.993}
\DefMacro{eval-bench-ewald-exec-tokyo-openmp-pkdb-min}{9.9784}
\DefMacro{eval-bench-ewald-exec-tokyo-openmp-pkdb-max}{481.993}
\DefMacro{eval-bench-ewald-exec-tokyo-openmp-avg-pkdb-slowdown}{1.406569}
\DefMacro{eval-bench-ewald-exec-tokyo-openmp-min-pkdb-slowdown}{1.006923}
\DefMacro{eval-bench-ewald-exec-tokyo-openmp-max-pkdb-slowdown}{4.579348}
\DefMacro{eval-bench-ewald-exec-tokyo-openmp-median-pkdb-slowdown}{1.086361}

\DefMacro{eval-bench-examinimd-exec-amd-mi300x-hip-pdb-y-1}{4.6151}
\DefMacro{eval-bench-examinimd-exec-amd-mi300x-hip-pdb-y-2}{7.6102}
\DefMacro{eval-bench-examinimd-exec-amd-mi300x-hip-pdb-y-3}{16.1492}
\DefMacro{eval-bench-examinimd-exec-amd-mi300x-hip-pdb-y-4}{33.2213}
\DefMacro{eval-bench-examinimd-exec-amd-mi300x-hip-pdb-y-5}{37.8217}
\DefMacro{eval-bench-examinimd-exec-amd-mi300x-hip-pdb-y-6}{40.1437}
\DefMacro{eval-bench-examinimd-exec-amd-mi300x-hip-pdb-y-7}{46.8248}
\DefMacro{eval-bench-examinimd-exec-amd-mi300x-hip-pdb-y-8}{48.2408}
\DefMacro{eval-bench-examinimd-exec-amd-mi300x-hip-pdb-y-9}{50.7727}
\DefMacro{eval-bench-examinimd-exec-amd-mi300x-hip-pdb-y-10}{54.2361}
\DefMacro{eval-bench-examinimd-exec-amd-mi300x-hip-pdb-y-11}{57.6071}
\DefMacro{eval-bench-examinimd-exec-amd-mi300x-hip-pdb-y-12}{60.9091}
\DefMacro{eval-bench-examinimd-exec-amd-mi300x-hip-pdb-y-13}{116.4342}
\DefMacro{eval-bench-examinimd-exec-amd-mi300x-hip-pdb-y-14}{232.1351}
\DefMacro{eval-bench-examinimd-exec-amd-mi300x-hip-pdb-y-15}{232.1351}
\DefMacro{eval-bench-examinimd-exec-amd-mi300x-hip-pdb-y-16}{581.7993}
\DefMacro{eval-bench-examinimd-exec-amd-mi300x-hip-pdb-min}{4.6151}
\DefMacro{eval-bench-examinimd-exec-amd-mi300x-hip-pdb-max}{581.7993}
\DefMacro{eval-bench-examinimd-exec-amd-mi300x-hip-pkdb-y-1}{10.6997}
\DefMacro{eval-bench-examinimd-exec-amd-mi300x-hip-pkdb-y-2}{14.8107}
\DefMacro{eval-bench-examinimd-exec-amd-mi300x-hip-pkdb-y-3}{26.3375}
\DefMacro{eval-bench-examinimd-exec-amd-mi300x-hip-pkdb-y-4}{49.377}
\DefMacro{eval-bench-examinimd-exec-amd-mi300x-hip-pkdb-y-5}{55.5583}
\DefMacro{eval-bench-examinimd-exec-amd-mi300x-hip-pkdb-y-6}{60.0045}
\DefMacro{eval-bench-examinimd-exec-amd-mi300x-hip-pkdb-y-7}{72.2306}
\DefMacro{eval-bench-examinimd-exec-amd-mi300x-hip-pkdb-y-8}{70.2554}
\DefMacro{eval-bench-examinimd-exec-amd-mi300x-hip-pkdb-y-9}{74.0087}
\DefMacro{eval-bench-examinimd-exec-amd-mi300x-hip-pkdb-y-10}{77.1451}
\DefMacro{eval-bench-examinimd-exec-amd-mi300x-hip-pkdb-y-11}{82.5272}
\DefMacro{eval-bench-examinimd-exec-amd-mi300x-hip-pkdb-y-12}{87.5607}
\DefMacro{eval-bench-examinimd-exec-amd-mi300x-hip-pkdb-y-13}{165.5425}
\DefMacro{eval-bench-examinimd-exec-amd-mi300x-hip-pkdb-y-14}{312.8967}
\DefMacro{eval-bench-examinimd-exec-amd-mi300x-hip-pkdb-y-15}{312.8967}
\DefMacro{eval-bench-examinimd-exec-amd-mi300x-hip-pkdb-y-16}{787.257}
\DefMacro{eval-bench-examinimd-exec-amd-mi300x-hip-pkdb-min}{10.6997}
\DefMacro{eval-bench-examinimd-exec-amd-mi300x-hip-pkdb-max}{787.257}
\DefMacro{eval-bench-examinimd-exec-amd-mi300x-hip-avg-pkdb-slowdown}{1.535331}
\DefMacro{eval-bench-examinimd-exec-amd-mi300x-hip-min-pkdb-slowdown}{1.347908}
\DefMacro{eval-bench-examinimd-exec-amd-mi300x-hip-max-pkdb-slowdown}{2.318411}
\DefMacro{eval-bench-examinimd-exec-amd-mi300x-hip-median-pkdb-slowdown}{1.456998}

\DefMacro{eval-bench-examinimd-exec-vista-gh200-cuda-pdb-y-1}{4.3716}
\DefMacro{eval-bench-examinimd-exec-vista-gh200-cuda-pdb-y-2}{7.5849}
\DefMacro{eval-bench-examinimd-exec-vista-gh200-cuda-pdb-y-3}{16.503}
\DefMacro{eval-bench-examinimd-exec-vista-gh200-cuda-pdb-y-4}{33.4675}
\DefMacro{eval-bench-examinimd-exec-vista-gh200-cuda-pdb-y-5}{37.9211}
\DefMacro{eval-bench-examinimd-exec-vista-gh200-cuda-pdb-y-6}{40.3162}
\DefMacro{eval-bench-examinimd-exec-vista-gh200-cuda-pdb-y-7}{46.0588}
\DefMacro{eval-bench-examinimd-exec-vista-gh200-cuda-pdb-y-8}{48.73}
\DefMacro{eval-bench-examinimd-exec-vista-gh200-cuda-pdb-y-9}{51.6724}
\DefMacro{eval-bench-examinimd-exec-vista-gh200-cuda-pdb-y-10}{55.0972}
\DefMacro{eval-bench-examinimd-exec-vista-gh200-cuda-pdb-y-11}{58.026}
\DefMacro{eval-bench-examinimd-exec-vista-gh200-cuda-pdb-y-12}{61.2898}
\DefMacro{eval-bench-examinimd-exec-vista-gh200-cuda-pdb-y-13}{117.5897}
\DefMacro{eval-bench-examinimd-exec-vista-gh200-cuda-pdb-y-14}{229.0653}
\DefMacro{eval-bench-examinimd-exec-vista-gh200-cuda-pdb-y-15}{570.5498}
\DefMacro{eval-bench-examinimd-exec-vista-gh200-cuda-pdb-min}{4.3716}
\DefMacro{eval-bench-examinimd-exec-vista-gh200-cuda-pdb-max}{570.5498}
\DefMacro{eval-bench-examinimd-exec-vista-gh200-cuda-pkdb-y-1}{19.7463}
\DefMacro{eval-bench-examinimd-exec-vista-gh200-cuda-pkdb-y-2}{22.8194}
\DefMacro{eval-bench-examinimd-exec-vista-gh200-cuda-pkdb-y-3}{33.8145}
\DefMacro{eval-bench-examinimd-exec-vista-gh200-cuda-pkdb-y-4}{54.4678}
\DefMacro{eval-bench-examinimd-exec-vista-gh200-cuda-pkdb-y-5}{60.0014}
\DefMacro{eval-bench-examinimd-exec-vista-gh200-cuda-pkdb-y-6}{63.3355}
\DefMacro{eval-bench-examinimd-exec-vista-gh200-cuda-pkdb-y-7}{69.9089}
\DefMacro{eval-bench-examinimd-exec-vista-gh200-cuda-pkdb-y-8}{73.6009}
\DefMacro{eval-bench-examinimd-exec-vista-gh200-cuda-pkdb-y-9}{76.1559}
\DefMacro{eval-bench-examinimd-exec-vista-gh200-cuda-pkdb-y-10}{80.4616}
\DefMacro{eval-bench-examinimd-exec-vista-gh200-cuda-pkdb-y-11}{84.268}
\DefMacro{eval-bench-examinimd-exec-vista-gh200-cuda-pkdb-y-12}{88.3799}
\DefMacro{eval-bench-examinimd-exec-vista-gh200-cuda-pkdb-y-13}{158.7279}
\DefMacro{eval-bench-examinimd-exec-vista-gh200-cuda-pkdb-y-14}{297.1531}
\DefMacro{eval-bench-examinimd-exec-vista-gh200-cuda-pkdb-y-15}{722.0926}
\DefMacro{eval-bench-examinimd-exec-vista-gh200-cuda-pkdb-min}{19.7463}
\DefMacro{eval-bench-examinimd-exec-vista-gh200-cuda-pkdb-max}{722.0926}
\DefMacro{eval-bench-examinimd-exec-vista-gh200-cuda-avg-pkdb-slowdown}{1.808301}
\DefMacro{eval-bench-examinimd-exec-vista-gh200-cuda-min-pkdb-slowdown}{1.265608}
\DefMacro{eval-bench-examinimd-exec-vista-gh200-cuda-max-pkdb-slowdown}{4.51695}
\DefMacro{eval-bench-examinimd-exec-vista-gh200-cuda-median-pkdb-slowdown}{1.510382}
\DefMacro{eval-bench-examinimd-exec-vista-gh200-openmp-pdb-y-1}{3.0872}
\DefMacro{eval-bench-examinimd-exec-vista-gh200-openmp-pdb-y-2}{6.1265}
\DefMacro{eval-bench-examinimd-exec-vista-gh200-openmp-pdb-y-3}{14.147}
\DefMacro{eval-bench-examinimd-exec-vista-gh200-openmp-pdb-y-4}{30.4294}
\DefMacro{eval-bench-examinimd-exec-vista-gh200-openmp-pdb-y-5}{34.3652}
\DefMacro{eval-bench-examinimd-exec-vista-gh200-openmp-pdb-y-6}{36.9271}
\DefMacro{eval-bench-examinimd-exec-vista-gh200-openmp-pdb-y-7}{42.2165}
\DefMacro{eval-bench-examinimd-exec-vista-gh200-openmp-pdb-y-8}{44.2785}
\DefMacro{eval-bench-examinimd-exec-vista-gh200-openmp-pdb-y-9}{47.1644}
\DefMacro{eval-bench-examinimd-exec-vista-gh200-openmp-pdb-y-10}{49.9849}
\DefMacro{eval-bench-examinimd-exec-vista-gh200-openmp-pdb-y-11}{53.1721}
\DefMacro{eval-bench-examinimd-exec-vista-gh200-openmp-pdb-y-12}{56.5551}
\DefMacro{eval-bench-examinimd-exec-vista-gh200-openmp-pdb-y-13}{111.1466}
\DefMacro{eval-bench-examinimd-exec-vista-gh200-openmp-pdb-y-14}{217.582}
\DefMacro{eval-bench-examinimd-exec-vista-gh200-openmp-pdb-y-15}{551.1709}
\DefMacro{eval-bench-examinimd-exec-vista-gh200-openmp-pdb-min}{3.0872}
\DefMacro{eval-bench-examinimd-exec-vista-gh200-openmp-pdb-max}{551.1709}
\DefMacro{eval-bench-examinimd-exec-vista-gh200-openmp-pkdb-y-1}{12.7115}
\DefMacro{eval-bench-examinimd-exec-vista-gh200-openmp-pkdb-y-2}{16.3125}
\DefMacro{eval-bench-examinimd-exec-vista-gh200-openmp-pkdb-y-3}{26.3436}
\DefMacro{eval-bench-examinimd-exec-vista-gh200-openmp-pkdb-y-4}{46.2581}
\DefMacro{eval-bench-examinimd-exec-vista-gh200-openmp-pkdb-y-5}{51.6888}
\DefMacro{eval-bench-examinimd-exec-vista-gh200-openmp-pkdb-y-6}{54.0083}
\DefMacro{eval-bench-examinimd-exec-vista-gh200-openmp-pkdb-y-7}{60.3147}
\DefMacro{eval-bench-examinimd-exec-vista-gh200-openmp-pkdb-y-8}{64.0106}
\DefMacro{eval-bench-examinimd-exec-vista-gh200-openmp-pkdb-y-9}{67.067}
\DefMacro{eval-bench-examinimd-exec-vista-gh200-openmp-pkdb-y-10}{71.4159}
\DefMacro{eval-bench-examinimd-exec-vista-gh200-openmp-pkdb-y-11}{75.2538}
\DefMacro{eval-bench-examinimd-exec-vista-gh200-openmp-pkdb-y-12}{78.6128}
\DefMacro{eval-bench-examinimd-exec-vista-gh200-openmp-pkdb-y-13}{148.4463}
\DefMacro{eval-bench-examinimd-exec-vista-gh200-openmp-pkdb-y-14}{275.9021}
\DefMacro{eval-bench-examinimd-exec-vista-gh200-openmp-pkdb-y-15}{693.215}
\DefMacro{eval-bench-examinimd-exec-vista-gh200-openmp-pkdb-min}{12.7115}
\DefMacro{eval-bench-examinimd-exec-vista-gh200-openmp-pkdb-max}{693.215}
\DefMacro{eval-bench-examinimd-exec-vista-gh200-openmp-avg-pkdb-slowdown}{1.701386}
\DefMacro{eval-bench-examinimd-exec-vista-gh200-openmp-min-pkdb-slowdown}{1.257713}
\DefMacro{eval-bench-examinimd-exec-vista-gh200-openmp-max-pkdb-slowdown}{4.117485}
\DefMacro{eval-bench-examinimd-exec-vista-gh200-openmp-median-pkdb-slowdown}{1.428749}

\DefMacro{eval-bench-examinimd-exec-tokyo-cuda-pdb-y-1}{4.2521}
\DefMacro{eval-bench-examinimd-exec-tokyo-cuda-pdb-y-2}{7.7909}
\DefMacro{eval-bench-examinimd-exec-tokyo-cuda-pdb-y-3}{17.3992}
\DefMacro{eval-bench-examinimd-exec-tokyo-cuda-pdb-y-4}{35.6384}
\DefMacro{eval-bench-examinimd-exec-tokyo-cuda-pdb-y-5}{39.7192}
\DefMacro{eval-bench-examinimd-exec-tokyo-cuda-pdb-y-6}{43.0913}
\DefMacro{eval-bench-examinimd-exec-tokyo-cuda-pdb-y-7}{47.4642}
\DefMacro{eval-bench-examinimd-exec-tokyo-cuda-pdb-y-8}{52.1771}
\DefMacro{eval-bench-examinimd-exec-tokyo-cuda-pdb-y-9}{54.3276}
\DefMacro{eval-bench-examinimd-exec-tokyo-cuda-pdb-y-10}{56.6203}
\DefMacro{eval-bench-examinimd-exec-tokyo-cuda-pdb-y-11}{61.1653}
\DefMacro{eval-bench-examinimd-exec-tokyo-cuda-pdb-y-12}{65.521}
\DefMacro{eval-bench-examinimd-exec-tokyo-cuda-pdb-y-13}{124.8138}
\DefMacro{eval-bench-examinimd-exec-tokyo-cuda-pdb-y-14}{243.2225}
\DefMacro{eval-bench-examinimd-exec-tokyo-cuda-pdb-y-15}{607.7113}
\DefMacro{eval-bench-examinimd-exec-tokyo-cuda-pdb-min}{4.2521}
\DefMacro{eval-bench-examinimd-exec-tokyo-cuda-pdb-max}{607.7113}
\DefMacro{eval-bench-examinimd-exec-tokyo-cuda-pkdb-y-1}{21.9871}
\DefMacro{eval-bench-examinimd-exec-tokyo-cuda-pkdb-y-2}{26.7729}
\DefMacro{eval-bench-examinimd-exec-tokyo-cuda-pkdb-y-3}{38.3591}
\DefMacro{eval-bench-examinimd-exec-tokyo-cuda-pkdb-y-4}{61.424}
\DefMacro{eval-bench-examinimd-exec-tokyo-cuda-pkdb-y-5}{66.4342}
\DefMacro{eval-bench-examinimd-exec-tokyo-cuda-pkdb-y-6}{70.6424}
\DefMacro{eval-bench-examinimd-exec-tokyo-cuda-pkdb-y-7}{77.8174}
\DefMacro{eval-bench-examinimd-exec-tokyo-cuda-pkdb-y-8}{81.3723}
\DefMacro{eval-bench-examinimd-exec-tokyo-cuda-pkdb-y-9}{85.7377}
\DefMacro{eval-bench-examinimd-exec-tokyo-cuda-pkdb-y-10}{89.0365}
\DefMacro{eval-bench-examinimd-exec-tokyo-cuda-pkdb-y-11}{93.2969}
\DefMacro{eval-bench-examinimd-exec-tokyo-cuda-pkdb-y-12}{97.9173}
\DefMacro{eval-bench-examinimd-exec-tokyo-cuda-pkdb-y-13}{173.7827}
\DefMacro{eval-bench-examinimd-exec-tokyo-cuda-pkdb-y-14}{322.7445}
\DefMacro{eval-bench-examinimd-exec-tokyo-cuda-pkdb-y-15}{782.8919}
\DefMacro{eval-bench-examinimd-exec-tokyo-cuda-pkdb-min}{21.9871}
\DefMacro{eval-bench-examinimd-exec-tokyo-cuda-pkdb-max}{782.8919}
\DefMacro{eval-bench-examinimd-exec-tokyo-cuda-avg-pkdb-slowdown}{1.948299}
\DefMacro{eval-bench-examinimd-exec-tokyo-cuda-min-pkdb-slowdown}{1.288263}
\DefMacro{eval-bench-examinimd-exec-tokyo-cuda-max-pkdb-slowdown}{5.17088}
\DefMacro{eval-bench-examinimd-exec-tokyo-cuda-median-pkdb-slowdown}{1.578161}
\DefMacro{eval-bench-examinimd-exec-tokyo-openmp-pdb-y-1}{4.5592}
\DefMacro{eval-bench-examinimd-exec-tokyo-openmp-pdb-y-2}{10.7708}
\DefMacro{eval-bench-examinimd-exec-tokyo-openmp-pdb-y-3}{25.8984}
\DefMacro{eval-bench-examinimd-exec-tokyo-openmp-pdb-y-4}{45.6133}
\DefMacro{eval-bench-examinimd-exec-tokyo-openmp-pdb-y-5}{57.2403}
\DefMacro{eval-bench-examinimd-exec-tokyo-openmp-pdb-y-6}{53.2151}
\DefMacro{eval-bench-examinimd-exec-tokyo-openmp-pdb-y-7}{60.5719}
\DefMacro{eval-bench-examinimd-exec-tokyo-openmp-pdb-y-8}{64.0634}
\DefMacro{eval-bench-examinimd-exec-tokyo-openmp-pdb-y-9}{68.5329}
\DefMacro{eval-bench-examinimd-exec-tokyo-openmp-pdb-y-10}{71.6522}
\DefMacro{eval-bench-examinimd-exec-tokyo-openmp-pdb-y-11}{75.479}
\DefMacro{eval-bench-examinimd-exec-tokyo-openmp-pdb-y-12}{88.2032}
\DefMacro{eval-bench-examinimd-exec-tokyo-openmp-pdb-y-13}{173.5313}
\DefMacro{eval-bench-examinimd-exec-tokyo-openmp-pdb-y-14}{322.5922}
\DefMacro{eval-bench-examinimd-exec-tokyo-openmp-pdb-y-15}{800.7376}
\DefMacro{eval-bench-examinimd-exec-tokyo-openmp-pdb-min}{4.5592}
\DefMacro{eval-bench-examinimd-exec-tokyo-openmp-pdb-max}{800.7376}
\DefMacro{eval-bench-examinimd-exec-tokyo-openmp-pkdb-y-1}{22.4227}
\DefMacro{eval-bench-examinimd-exec-tokyo-openmp-pkdb-y-2}{30.0503}
\DefMacro{eval-bench-examinimd-exec-tokyo-openmp-pkdb-y-3}{47.3612}
\DefMacro{eval-bench-examinimd-exec-tokyo-openmp-pkdb-y-4}{71.1017}
\DefMacro{eval-bench-examinimd-exec-tokyo-openmp-pkdb-y-5}{77.0921}
\DefMacro{eval-bench-examinimd-exec-tokyo-openmp-pkdb-y-6}{81.3208}
\DefMacro{eval-bench-examinimd-exec-tokyo-openmp-pkdb-y-7}{89.4599}
\DefMacro{eval-bench-examinimd-exec-tokyo-openmp-pkdb-y-8}{93.8249}
\DefMacro{eval-bench-examinimd-exec-tokyo-openmp-pkdb-y-9}{97.3951}
\DefMacro{eval-bench-examinimd-exec-tokyo-openmp-pkdb-y-10}{108.4381}
\DefMacro{eval-bench-examinimd-exec-tokyo-openmp-pkdb-y-11}{110.9898}
\DefMacro{eval-bench-examinimd-exec-tokyo-openmp-pkdb-y-12}{111.7348}
\DefMacro{eval-bench-examinimd-exec-tokyo-openmp-pkdb-y-13}{207.6739}
\DefMacro{eval-bench-examinimd-exec-tokyo-openmp-pkdb-y-14}{390.6939}
\DefMacro{eval-bench-examinimd-exec-tokyo-openmp-pkdb-y-15}{952.0208}
\DefMacro{eval-bench-examinimd-exec-tokyo-openmp-pkdb-min}{22.4227}
\DefMacro{eval-bench-examinimd-exec-tokyo-openmp-pkdb-max}{952.0208}
\DefMacro{eval-bench-examinimd-exec-tokyo-openmp-avg-pkdb-slowdown}{1.745378}
\DefMacro{eval-bench-examinimd-exec-tokyo-openmp-min-pkdb-slowdown}{1.18893}
\DefMacro{eval-bench-examinimd-exec-tokyo-openmp-max-pkdb-slowdown}{4.918122}
\DefMacro{eval-bench-examinimd-exec-tokyo-openmp-median-pkdb-slowdown}{1.470473}

\DefMacro{eval-debugtime-tokyo-baseline-t-y-1}{152.9127}
\DefMacro{eval-debugtime-tokyo-baseline-t-y-2}{316.7578}
\DefMacro{eval-debugtime-tokyo-baseline-t-y-3}{315.3221}
\DefMacro{eval-debugtime-tokyo-baseline-t-y-4}{685.8267}
\DefMacro{eval-debugtime-tokyo-baseline-t-y-5}{666.5894}
\DefMacro{eval-debugtime-tokyo-baseline-t-min}{152.9127}
\DefMacro{eval-debugtime-tokyo-baseline-t-max}{685.8267}
\DefMacro{eval-debugtime-tokyo-pkdbcuda-t-y-1}{5.7515}
\DefMacro{eval-debugtime-tokyo-pkdbcuda-t-y-2}{5.6232}
\DefMacro{eval-debugtime-tokyo-pkdbcuda-t-y-3}{5.5253}
\DefMacro{eval-debugtime-tokyo-pkdbcuda-t-y-4}{5.7185}
\DefMacro{eval-debugtime-tokyo-pkdbcuda-t-y-5}{5.9289}
\DefMacro{eval-debugtime-tokyo-pkdbcuda-t-min}{5.5253}
\DefMacro{eval-debugtime-tokyo-pkdbcuda-t-max}{5.9289}
\DefMacro{eval-debugtime-tokyo-pkdbopenmp-t-y-1}{6.8117}
\DefMacro{eval-debugtime-tokyo-pkdbopenmp-t-y-2}{7.0743}
\DefMacro{eval-debugtime-tokyo-pkdbopenmp-t-y-3}{7.2087}
\DefMacro{eval-debugtime-tokyo-pkdbopenmp-t-y-4}{9.8863}
\DefMacro{eval-debugtime-tokyo-pkdbopenmp-t-y-5}{7.626}
\DefMacro{eval-debugtime-tokyo-pkdbopenmp-t-min}{6.8117}
\DefMacro{eval-debugtime-tokyo-pkdbopenmp-t-max}{9.8863}

\DefMacro{eval-debugtime-vista-gh200-baseline-t-y-1}{94.8349}
\DefMacro{eval-debugtime-vista-gh200-baseline-t-y-2}{194.4358}
\DefMacro{eval-debugtime-vista-gh200-baseline-t-y-3}{195.1492}
\DefMacro{eval-debugtime-vista-gh200-baseline-t-y-4}{416.0527}
\DefMacro{eval-debugtime-vista-gh200-baseline-t-y-5}{417.8}
\DefMacro{eval-debugtime-vista-gh200-baseline-t-min}{94.8349}
\DefMacro{eval-debugtime-vista-gh200-baseline-t-max}{417.8}
\DefMacro{eval-debugtime-vista-gh200-pkdbcuda-t-y-1}{19.6148}
\DefMacro{eval-debugtime-vista-gh200-pkdbcuda-t-y-2}{18.2408}
\DefMacro{eval-debugtime-vista-gh200-pkdbcuda-t-y-3}{18.026}
\DefMacro{eval-debugtime-vista-gh200-pkdbcuda-t-y-4}{18.7166}
\DefMacro{eval-debugtime-vista-gh200-pkdbcuda-t-y-5}{19}
\DefMacro{eval-debugtime-vista-gh200-pkdbcuda-t-min}{18.026}
\DefMacro{eval-debugtime-vista-gh200-pkdbcuda-t-max}{19.6148}
\DefMacro{eval-debugtime-vista-gh200-pkdbopenmp-t-y-1}{12.0484}
\DefMacro{eval-debugtime-vista-gh200-pkdbopenmp-t-y-2}{12.0159}
\DefMacro{eval-debugtime-vista-gh200-pkdbopenmp-t-y-3}{12.0758}
\DefMacro{eval-debugtime-vista-gh200-pkdbopenmp-t-y-4}{12.1358}
\DefMacro{eval-debugtime-vista-gh200-pkdbopenmp-t-y-5}{12.2}
\DefMacro{eval-debugtime-vista-gh200-pkdbopenmp-t-min}{12.0159}
\DefMacro{eval-debugtime-vista-gh200-pkdbopenmp-t-max}{12.2}

\DefMacro{eval-debugtime-amd-mi300x-baseline-t-y-1}{110.125}
\DefMacro{eval-debugtime-amd-mi300x-baseline-t-y-2}{225.0134}
\DefMacro{eval-debugtime-amd-mi300x-baseline-t-y-3}{225.8787}
\DefMacro{eval-debugtime-amd-mi300x-baseline-t-y-4}{486.0081}
\DefMacro{eval-debugtime-amd-mi300x-baseline-t-y-5}{480.179}
\DefMacro{eval-debugtime-amd-mi300x-baseline-t-min}{110.125}
\DefMacro{eval-debugtime-amd-mi300x-baseline-t-max}{486.0081}
\DefMacro{eval-debugtime-amd-mi300x-pkdbhip-t-y-1}{10.3596}
\DefMacro{eval-debugtime-amd-mi300x-pkdbhip-t-y-2}{10.3693}
\DefMacro{eval-debugtime-amd-mi300x-pkdbhip-t-y-3}{10.3823}
\DefMacro{eval-debugtime-amd-mi300x-pkdbhip-t-y-4}{10.4555}
\DefMacro{eval-debugtime-amd-mi300x-pkdbhip-t-y-5}{10.4548}
\DefMacro{eval-debugtime-amd-mi300x-pkdbhip-t-min}{10.3596}
\DefMacro{eval-debugtime-amd-mi300x-pkdbhip-t-max}{10.4555}
}{}
\IfFileExists{tables/eval-examinimd-breakdown-macros.tex}{
\DefMacro{eval-examinimd-breakdown-4000-startup}{3.0}
\DefMacro{eval-examinimd-breakdown-4000-bdb_line_tracing}{58.8}
\DefMacro{eval-examinimd-breakdown-4000-workunit_dispatch}{0.0}
\DefMacro{eval-examinimd-breakdown-4000-pykokkos_execution}{27.5}
\DefMacro{eval-examinimd-breakdown-4000-kernel_exec}{13.7}
\DefMacro{eval-examinimd-breakdown-4000-app_execution}{6.0}

\DefMacro{eval-examinimd-breakdown-5000-startup}{2.3}
\DefMacro{eval-examinimd-breakdown-5000-bdb_line_tracing}{58.4}
\DefMacro{eval-examinimd-breakdown-5000-workunit_dispatch}{0.0}
\DefMacro{eval-examinimd-breakdown-5000-pykokkos_execution}{28.1}
\DefMacro{eval-examinimd-breakdown-5000-kernel_exec}{13.7}
\DefMacro{eval-examinimd-breakdown-5000-app_execution}{6.6}

\DefMacro{eval-examinimd-breakdown-10000-startup}{2.4}
\DefMacro{eval-examinimd-breakdown-10000-bdb_line_tracing}{57.6}
\DefMacro{eval-examinimd-breakdown-10000-workunit_dispatch}{0.0}
\DefMacro{eval-examinimd-breakdown-10000-pykokkos_execution}{27.0}
\DefMacro{eval-examinimd-breakdown-10000-kernel_exec}{12.4}
\DefMacro{eval-examinimd-breakdown-10000-app_execution}{9.4}

\DefMacro{eval-examinimd-breakdown-20000-startup}{2.1}
\DefMacro{eval-examinimd-breakdown-20000-bdb_line_tracing}{53.6}
\DefMacro{eval-examinimd-breakdown-20000-workunit_dispatch}{0.0}
\DefMacro{eval-examinimd-breakdown-20000-pykokkos_execution}{25.6}
\DefMacro{eval-examinimd-breakdown-20000-kernel_exec}{14.3}
\DefMacro{eval-examinimd-breakdown-20000-app_execution}{12.8}

\DefMacro{eval-examinimd-breakdown-32000-startup}{2.0}
\DefMacro{eval-examinimd-breakdown-32000-bdb_line_tracing}{50.7}
\DefMacro{eval-examinimd-breakdown-32000-workunit_dispatch}{0.0}
\DefMacro{eval-examinimd-breakdown-32000-pykokkos_execution}{23.7}
\DefMacro{eval-examinimd-breakdown-32000-kernel_exec}{13.3}
\DefMacro{eval-examinimd-breakdown-32000-app_execution}{18.0}

\DefMacro{eval-examinimd-breakdown-108000-startup}{1.4}
\DefMacro{eval-examinimd-breakdown-108000-bdb_line_tracing}{36.0}
\DefMacro{eval-examinimd-breakdown-108000-workunit_dispatch}{0.0}
\DefMacro{eval-examinimd-breakdown-108000-pykokkos_execution}{17.2}
\DefMacro{eval-examinimd-breakdown-108000-kernel_exec}{14.4}
\DefMacro{eval-examinimd-breakdown-108000-app_execution}{36.7}

\DefMacro{eval-examinimd-breakdown-256000-startup}{0.9}
\DefMacro{eval-examinimd-breakdown-256000-bdb_line_tracing}{26.0}
\DefMacro{eval-examinimd-breakdown-256000-workunit_dispatch}{0.0}
\DefMacro{eval-examinimd-breakdown-256000-pykokkos_execution}{14.4}
\DefMacro{eval-examinimd-breakdown-256000-kernel_exec}{13.5}
\DefMacro{eval-examinimd-breakdown-256000-app_execution}{52.2}

\DefMacro{eval-examinimd-breakdown-300000-startup}{0.8}
\DefMacro{eval-examinimd-breakdown-300000-bdb_line_tracing}{21.1}
\DefMacro{eval-examinimd-breakdown-300000-workunit_dispatch}{0.0}
\DefMacro{eval-examinimd-breakdown-300000-pykokkos_execution}{10.1}
\DefMacro{eval-examinimd-breakdown-300000-kernel_exec}{14.1}
\DefMacro{eval-examinimd-breakdown-300000-app_execution}{57.2}

\DefMacro{eval-examinimd-breakdown-320000-startup}{0.8}
\DefMacro{eval-examinimd-breakdown-320000-bdb_line_tracing}{20.3}
\DefMacro{eval-examinimd-breakdown-320000-workunit_dispatch}{0.0}
\DefMacro{eval-examinimd-breakdown-320000-pykokkos_execution}{9.7}
\DefMacro{eval-examinimd-breakdown-320000-kernel_exec}{14.1}
\DefMacro{eval-examinimd-breakdown-320000-app_execution}{58.3}

\DefMacro{eval-examinimd-breakdown-370000-startup}{0.8}
\DefMacro{eval-examinimd-breakdown-370000-bdb_line_tracing}{18.5}
\DefMacro{eval-examinimd-breakdown-370000-workunit_dispatch}{0.0}
\DefMacro{eval-examinimd-breakdown-370000-pykokkos_execution}{8.7}
\DefMacro{eval-examinimd-breakdown-370000-kernel_exec}{13.7}
\DefMacro{eval-examinimd-breakdown-370000-app_execution}{61.2}
}{}

\begin{document}

\title{Interactive Debugger for \\ Performance Portable Python HPC Kernels}

\author{
      \IEEEauthorblockN{Ivan Grigorik}
      \IEEEauthorblockA{The University of Texas at Austin \\
            Austin, Texas, USA \\
            grigorik@utexas.edu}
      \and
      \IEEEauthorblockN{Gabriel Kosmacher}
      \IEEEauthorblockA{The University of Texas at Austin \\
            Austin, Texas, USA \\
            gkosmacher@utexas.edu}
      \linebreakand
      \IEEEauthorblockN{George Biros}
      \IEEEauthorblockA{The University of Texas at Austin \\
            Austin, Texas, USA \\
            gbiros@acm.org}
      \and
      \IEEEauthorblockN{Milos Gligoric}
      \IEEEauthorblockA{The University of Texas at Austin \\
            Austin, Texas, USA \\
            gligoric@utexas.edu}
}

\maketitle
\thispagestyle{fancy}
\lhead{}
\rhead{}
\chead{}
\lfoot{\footnotesize{
            SC26, November 15-20, 2026, Chicago, Illinois, USA
            \newline 979-8-3195-4789-7/26/\$31.00 \copyright 2026 IEEE}}
\rfoot{}
\cfoot{}
\renewcommand{\headrulewidth}{0pt}
\renewcommand{\footrulewidth}{0pt}

\setcounter{page}{1}

\begin{abstract}
      We propose \ourTool, the first interactive debugger for GPU and
      multithreaded low-level kernels written in \Python.  \Python is widely
      used in high performance computing (\HPC), with frameworks such as
      \PyKokkos translating \Python-embedded domain-specific languages to
      native code that runs across \OpenMP-threaded \CPUs and various \GPUs.
      Yet interactive debugging support for such code is absent: developers
      resort to print statements, framework-specific assertions, or CPU-only
      execution,
      the last of which requires altering the program or its data and can
      mask device-specific bugs.
      \ourTool enables standard interactive debugging like breakpoints, stepping,
      and variable inspection while preserving actual on-device execution
      without source modification.
      Beyond these fundamentals, \ourTool introduces two advanced
      capabilities that exploit the dynamic nature of \Python and \PyKokkos:
      (i)~\Liveeval, which lets developers execute arbitrary \Python
      expressions or entire kernels in the middle of a paused kernel without
      restarting the process;
      (ii)~\KernelSubstitute, which allows an actively running kernel to be updated
      and reloaded on the fly, so only the kernel is recompiled and
      re-executed without restarting the application.
      Our performance evaluation on \Intel, \AMD, and \NVIDIA \CPUs and \NVIDIA and
      \AMD \GPUs shows that \ourTool introduces limited overhead and is
      practical for everyday use while introducing critical debugging features
      to the \Python \HPC ecosystem.
\end{abstract}

\begin{IEEEkeywords}
      High Performance Computing, Python, Debugging, Performance Portability.
\end{IEEEkeywords}

\section{Introduction}

The rapid diversification of high performance computing (\HPC)
hardware platforms, from \OpenMP-threaded \CPUs to \GPUs and
specialized accelerators, makes it increasingly challenging to develop
\HPC applications that run efficiently across different devices.
Performance-portable frameworks such as Kokkos~\cite{kokkos} address
this by enabling developers to write code once and deploy it across
different architectures without manual porting.  While such frameworks
were traditionally implemented in statically typed languages like C++,
the growing need for rapid prototyping and seamless integration with
machine learning workflows
has driven the development of equivalent \Python-based frameworks.
These frameworks compile \Python kernels just-in-time to C++ or other
lower-level formats; \PyKokkos~\cite{PyKokkos}, for instance,
translates \Python-embedded domain-specific language (\eDSL) kernels
directly to Kokkos C++, enabling developers to harness
performance-portable \HPC without leaving the \Python ecosystem.

Despite substantial progress on frameworks covering new hardware
backends, language features, and compilation pipelines, there has been
minimal effort to provide tools that \emph{assist the development process}
itself.
We designed and developed \ourTool, the first interactive debugger for
GPU and multithreaded low-level kernels written in \Python, targeting
\PyKokkos kernels in particular.

The current state-of-practice for debugging \Python \eDSL{}s is embedded
\CodeIn{print} statements~\cite{pyCudaDebugging,numbaCudaDebug} or
assertions~\cite{TritonLang, debugHPC, KComps}; a small subset of the
debugging capabilities available to device-native kernel developers
provided by \CUDAGDB~\cite{cudagdbDocumentation} or
\ROCGDB~\cite{ROCgdbDocumentation}.
As an alternative, developers may choose to execute the program serially
on the \CPU without translation~\cite{PyKokkos, tritonDebugging} or with
partial translation; for complex workloads, this approach forces the developer to alter the program or
its data to keep execution time manageable and may fail to expose
corrupted behavior specific to parallel execution on the target
device.
Building a proper interactive debugger for this setting is itself
non-trivial: it must bridge the \Python host with JIT-compiled device
kernels, map Python-level source concepts to dynamically generated
device code, and unify distinct low-level debugging protocols for
\GPU (\CUDAGDB/\ROCGDB) and \OpenMP-threaded \CPU targets.

\ourTool is designed to be fully familiar to \Python developers by mirroring the
interface of \PDB~\cite{python_pdb_module}, integrating naturally into existing
development workflows. At the same time, its architecture supports debugging
application kernels executing on a device (e.g., a \GPU), preserving actual
on-device execution and performance, without source modification.

Beyond standard interactive debugging features such as breakpoints,
stepping, continuation, and variable inspection, \ourTool introduces two
novel capabilities that exploit the dynamic nature of \Python and
\PyKokkos:
\begin{enumerate*}
      \item {\emph{\Liveeval}, which lets developers execute arbitrary \Python expressions
            or entire kernels
            in the middle of a paused kernel without restarting the process, enabling
            inspection or modification of program state at any point during debugging.
            To reduce the latency of live evaluation, \ourTool can execute multiple
            kernel versions concurrently rather than sequentially.}
      \item {\emph{\KernelSubstitute}, which allows an actively running kernel to be updated and
            reloaded on the fly, so only the kernel is recompiled and re-executed
            without restarting the entire application.}
\end{enumerate*}
Due to its modular design, \ourTool currently supports \Intel, \AMD and \NVIDIA
\CPUs and \NVIDIA and \AMD \GPUs; adding a new platform requires only
implementing a platform-specific protocol without changes to the rest of
the framework.

\noindent
The key contributions of this paper include:
\begin{itemize}[topsep=0pt,itemsep=1pt,partopsep=0ex,parsep=0ex,leftmargin=*]

      \item[$\star$] \textbf{Conceptualization.} We present \ourTool, the first
            interactive debugger for GPU and multithreaded low-level \Python
            kernels.  \ourTool enables on-device debugging without source
            modification and integrates seamlessly into the familiar
            \CodeIn{pdb} interface, supporting context switching between \Python
            and generated device code, and across heterogeneous accelerator
            types (e.g., \CUDA and \OpenMP), within a single session.

      \item[$\star$] \textbf{\Liveeval.} \ourTool enables execution of
            arbitrary code---on the \Python host or directly on the
            device---in the middle of a paused kernel session, giving
            developers the ability to inspect and modify program state at any
            point without restarting the application.

      \item[$\star$] \textbf{\KernelSubstitute.} \ourTool supports in-flight
            kernel changes, allowing developers to fix and continue kernel
            execution within the same debugger session without a full
            application restart.

      \item[$\star$] \textbf{Performance evaluation.} We evaluate \ourTool on
            representative \PyKokkos workloads: ExaMiniMD,
            a Boltzmann-kinetics solver, and a periodic Ewald sum for the Stokes
            potential.  We quantify end-to-end debugging over \ourTool supported on \Intel, \NVIDIA
            and \AMD \CPUs and \NVIDIA and \AMD \GPUs, compare against baseline
            debugging workflows, and measure the time saved by call site
            substitution versus a full application restart.

\end{itemize}

\vspace{3pt}
\noindent
In doing so, \ourTool brings long-overdue, first-class debugging support to
developers of performance-portable \Python \HPC kernels.
\ourTool is publicly available at \ourToolGitGub.

\section{Background and Example}

In this section, we provide a brief description of \Kokkos
(\S\ref{KokkosBackground}) and \PyKokkos
(\S\ref{PyKokkosBackground}), as well as showcase an example of a
debugging session using \ourTool (\S\ref{pkdbExample}).

\subsection{The \Kokkos framework}
\label{KokkosBackground}

\Kokkos~\cite{KokkosEcosystem2021} is a C++ framework for developing
performance portable \HPC applications.  It is implemented as a
heavily templated, header-only library that maps a single source to
diverse hardware backends, e.g., \Intel, \AMD and \NVIDIA \CPUs
(\CodeIn{\OpenMP}, \CodeIn{Threads}, \CodeIn{Serial}) and \NVIDIA and \AMD \GPUs
(\CodeIn{\CUDA}, \CodeIn{\HIP}).

Kernels in \Kokkos are expressed as C++ functors or lambdas and launched via one
of three \emph{parallel dispatch} operations: \CodeIn{parallel\_for} (map),
\CodeIn{parallel\_reduce} (reduction), and \CodeIn{parallel\_scan} (prefix
scan). Each dispatch is parameterized by two key abstractions: executions spaces
and execution policies. An \emph{execution space} (e.g., \CodeIn{CUDA},
\CodeIn{HIP}, \CodeIn{OpenMP}) selects the target device and its associated API.
An \emph{execution policy} governs how work is mapped onto that device:
\CodeIn{RangePolicy} specifies a one-dimensional iteration range whose
iterations are distributed across threads, while \CodeIn{TeamPolicy} introduces
hierarchical parallelism by grouping threads into \emph{teams} (analogous to
\CUDA thread blocks) that can synchronize and share local scratch memory. Data
is managed through the \CodeIn{View} data structure, a multi-dimensional array
abstraction whose memory spaces (which are logically distinct physical memory
resources where data is stored) and layout are parameterized to match the target
backend.

Despite its portability, \Kokkos presents significant usability
challenges.
The heavy reliance on C++ template metaprogramming makes compiler
diagnostics verbose and difficult to interpret, and explicit
management of memory spaces and device placement adds overhead to the
development workflow.
Furthermore, C++ is a poor fit for developers who work primarily in
\Python---whether for rapid prototyping, data analysis, or integration with
machine learning frameworks and libraries---creating a barrier between
\HPC kernel development and the broader scientific computing ecosystem.

\lstset{
  basicstyle=\ttfamily\scriptsize,
  keywordstyle=\bfseries,
  commentstyle=\itshape\color{gray},
  numbers=left,
  numberstyle=\ttfamily\scriptsize,
  numbersep=0.9em,
  showstringspaces=false,
  flexiblecolumns=false,
  breaklines=true,
  frame=single,
  framerule=0.4pt,
  xleftmargin=1em,
  xrightmargin=1em,
  aboveskip=2pt,
  belowskip=2pt,
}

\begin{figure}[t]
  \centering
  \begin{subfigure}{\linewidth}
    \lstset{language=Python}
    \begin{lstlisting}[language=python, escapechar=|]
import cupy as cp
import pykokkos as pk

|\colorbox{green!22}{\texttt{@pk.workunit}}|   |\textcolor{ForestGreen}{\textbf{\PyKokkos \HPC kernel executed on a \GPU}}||\label{yax:wokunit}|
def yAx(j, acc, cols, y_view, x_view, A_view):    |\label{line:lineYAX}|
    temp2: float = 0                              |\label{line:pyTemp2LOC}|
    for i in range(cols):                         |\label{line:pyForLOC}|
      temp2 += A_view[j][i] * x_view[i]             |\label{line:pyTempSumLOC}|
    acc += y_view[j] * temp2                      |\label{line:pyAccLOC}|

def run() -> None:
    N = 256   # Rows
    M = 1024  # Cols
    y = cp.random.rand(N)                |\label{line:inityLOC}|
    x = cp.random.rand(M)                |\label{line:initxLOC}|
    A = cp.random.rand(N, M)             |\label{line:initALOC}|
            |\tikz[baseline=(ann.base)] \node[draw=green, dashed, rounded corners=1pt, inner xsep=3pt, inner ysep=1pt, text=ForestGreen, font=\bfseries\scriptsize] (ann) {Execution space set to be a GPU device};|
    space = pk.RangePolicy(|\colorbox{green!22}{\texttt{pk.ExecutionSpace.Cuda}}|, 0,N)|\label{line:pyPolicyLine}|
    pk.parallel_reduce(space, yAx, cols=M, |\label{line:reduceLine}|
                       y_view=y, x_view=x, A_view=A)

if __name__ == "__main__":
    run()
    \end{lstlisting}
    \caption{Python code with~\CodeIn{PyKokkos} kernel (\CodeIn{yax.py}).}
    \label{fig:yax-python}
  \end{subfigure}

  \vspace{0.75em}

  \begin{subfigure}{\linewidth}
    \begin{lstlisting}[language=C++, escapechar=|]
KOKKOS_FUNCTION void operator()(const yAx_tag &,  
                     int32_t j, double &acc) const {
  double temp2 = 0;                             |\label{cppTemp2LOC}|
  for (int32_t i = 0; i < cols; i += 1)        |\label{cppForLOC}|
    temp2 += A_view(j, i) * x_view(i);          |\label{cppTempSumLOC}|
  acc += y_view(j) * temp2; };                    |\label{cppAccLOC}|
    \end{lstlisting}
    \caption{Generated C++ functor from~\CodeIn{yAx} workunit.}
    \label{fig:yax-cpp}
  \end{subfigure}
  \caption{End-to-end~\PyKokkos example. \ref{fig:yax-python} shows an
  HPC kernel (function annotated with \CodeIn{@pk.workunit}, lines~\ref{line:lineYAX}--\ref{line:pyAccLOC}) and the
  code that launches said kernel.  All code without the kernel is
  executed by the standard Python interpreter, while the kernel is
  automatically translated (at the invocation time) to \Kokkos (\ref{fig:yax-cpp}) and executed
  using the targeted execution space (\CUDA in this
  example).\label{fig:yax-example}}
\end{figure}

\subsection{\PyKokkos}
\label{PyKokkosBackground}

\PyKokkos~\cite{PyKokkos} was developed to remove the limitations of \Kokkos for
\Python developers while preserving its performance portability.  Developers
write \HPC kernels in a subset of \Python (\eDSL) and \PyKokkos automatically
translates, compiles, and executes them on the target device with minimal
overhead.  The \eDSL mirrors the \Kokkos dispatch API: kernels are functions
decorated with \CodeIn{@pk.workunit} (Figure~\ref{fig:yax-example},
line~\ref{yax:wokunit}) and later launched via \CodeIn{parallel\_for},
\CodeIn{parallel\_reduce}, or \CodeIn{parallel\_scan} and paired with execution
spaces and policies as in \Kokkos.  Crucially, \PyKokkos does not introduce any
extra data abstractions: rather than requiring developers to manage \Kokkos
\CodeIn{View}s explicitly, it accepts standard \Python array
API~\cite{python_array_api_2023} objects directly, e.g., NumPy and CuPy arrays, as well as
PyTorch tensors. \PyKokkos then converts these objects to \CodeIn{View}s
internally with minimal overhead; keeping the \Python side of the code idiomatic
and compatible with the broader scientific and machine learning ecosystem.
Because \PyKokkos kernels remain ordinary \Python functions at runtime, they
retain the dynamic properties of \Python---introspection, runtime code
modification, and late binding---on which \ourTool is built.

Figure~\ref{fig:yax-example} illustrates the \PyKokkos workflow through a fused
vector-matrix-vector operation $\boldsymbol{y}\boldsymbol{A}\boldsymbol{x}$,
where $\boldsymbol{y}\in \mathbb{R}^N$, $\boldsymbol{x}\in \mathbb{R}^M$ and $\boldsymbol{A} \in
      \mathbb{R}^{N \times M}$.  The input arrays are \CuPy arrays instantiated on the
\GPU (lines \ref{line:inityLOC}--\ref{line:initALOC}).  The kernel \CodeIn{yAx} (line
\ref{line:lineYAX}) is decorated with \CodeIn{@pk.workunit} and launched by a
\CodeIn{parallel\_reduce} over a \CodeIn{RangePolicy} of $N$ threads on the
\CodeIn{\CUDA} execution space (lines \ref{line:pyPolicyLine},~\ref{line:reduceLine}).  \PyKokkos
translates \CodeIn{yAx} into the \Kokkos C++ functor shown in
Figure~\ref{fig:yax-cpp}, compiles it just-in-time, and executes it on the \GPU;
the results are written back through pointers to the original \Python objects
and remain accessible to the rest of the \Python runtime.

\subsection{\ourTool example}
\label{pkdbExample}

\lstset{
  basicstyle=\ttfamily\scriptsize,
  keywordstyle=\bfseries,
  commentstyle=\itshape\color{gray},
  numbers=left,
  numberstyle=\ttfamily\scriptsize,
  numbersep=0.9em,
  showstringspaces=false,
  flexiblecolumns=false,
  breaklines=true,
  frame=single,
  framerule=0.4pt,
  xleftmargin=2em,
  xrightmargin=1em,
  aboveskip=2pt,
  belowskip=2pt,
}

\providecommand{\pkdbKeyLineBg}{%
  \leavevmode\rlap{%
    \setlength{\fboxsep}{0pt}%
    \colorbox{yellow!35}{\makebox[\linewidth][l]{\strut}}%
  }%
}
\providecommand{\pkdbDeviceHostMatchBg}{%
  \leavevmode\rlap{%
    \setlength{\fboxsep}{0pt}%
    \colorbox{green!22}{\makebox[\linewidth][l]{\strut}}%
  }%
}

\begin{figure}[t]
  \centering
  \begin{lstlisting}[
    language={},
    escapechar=|,
    literate=
      {(pkdb-cuda)}{{\textcolor{ForestGreen!85!black}{\bfseries(pkdb-cuda)}}}{11}
      {(pkdb)}{{\textcolor{MidnightBlue!70!black}{\bfseries(pkdb)}}}{6}
  ]
(pkdb) break |\ref{line:pyTemp2LOC}| |\ref{line:pyAccLOC}| |\ref{line:pyPolicyLine}| |\label{exp:setBreak}|
Breakpoint set at yax.py:|\ref{line:pyTemp2LOC}| |\ref{line:pyAccLOC}| |\ref{line:pyPolicyLine}|
(pkdb) run|\label{exp:run}|
> /root/yax.py(|\ref{line:pyPolicyLine}|)run()|\label{exp:pyPolicyLineHit}|
-> space = pk.RangePolicy(pk.ExecutionSpace.Cuda, 0,N)
(pkdb) print y[0:3]|\label{exp:PyPrintY}|
array([0.81069483, 0.0676661 , 0.51309116])
(pkdb) continue|\label{exp:PyContinue}|
CUDA kernel breakpoint hit at yax.py:|\ref{line:pyTemp2LOC}||\label{exp:kerneLineHit1}|
(pkdb-cuda) where|\label{exp:kernelWhere}|
#0 /root/yax.py:|\ref{line:pyTemp2LOC}| (yAx)
#1 /root/yax.py:|\ref{line:pyPolicyLine}| (main)
(pkdb-cuda) print y_view[0:2]|\label{exp:kernelPrintY02}|
list([0.810694828106066, 0.06766610250020871])
|\pkdbKeyLineBg|(pkdb-cuda) print j|\label{exp:kernelPrintJ0}|
|\pkdbKeyLineBg|int(0)
|\pkdbKeyLineBg|(pkdb-cuda) print y_view[j:j+2]|\label{exp:kernelPrintYjj2}|
|\pkdbKeyLineBg|list([0.810694828106066, 0.06766610250020871])
(pkdb-cuda) continue|\label{exp:CppContinue}|
CUDA kernel breakpoint hit at yax.py:|\ref{line:pyAccLOC}||\label{exp:kernelLineHit2}|
|\pkdbDeviceHostMatchBg|(pkdb-cuda) eval device_sum(y_view)|\label{exp:kernelDeviceSum}|
|\pkdbDeviceHostMatchBg|float(134.55991968052976)
|\pkdbKeyLineBg|(pkdb-cuda) info cuda threads|\label{exp:kernelInfoCudaThreads}|
|\pkdbKeyLineBg|* block_idx=(0,0,0)  thread_idx=(0,0,0)..(0,255,0)  to_block_idx=(0,0,0)  count=256
|\pkdbKeyLineBg|(pkdb-cuda) cuda thread 0 10|\label{exp:kernelCudaThread010}|
|\pkdbKeyLineBg|(pkdb-cuda) print j|\label{exp:kernelPrintJ10}|
|\pkdbKeyLineBg|int(10)
(pkdb-cuda) finish|\label{exp:CppContinue2}|
% > /root/yax.py()run()|\label{exp:pyJsonLineHit}|
|\pkdbDeviceHostMatchBg|(pkdb) sum(y)|\label{exp:pySumY}|
|\pkdbDeviceHostMatchBg|array(134.55991968)
  \end{lstlisting}
  \caption{Example \ourTool debugging session on~\CodeIn{yax.py} (from Figure~\ref{fig:yax-example}).  \label{fig:debug-experience}}
  
\end{figure}

Figure~\ref{fig:debug-experience} illustrates a \ourTool debugging session on
the \CodeIn{yAx} program from Figure~\ref{fig:yax-example}. The developer sets
breakpoints (line~\ref{exp:setBreak}) by \Python source line, either inside the
\CodeIn{@pk.workunit} kernel or in the outer \Python code, using a single,
uniform interface. Internally, \ourTool dispatches each command to the
appropriate debugger: \CUDAGDB handles breakpoints inside the device kernel, and
\PDB handles breakpoints in the \Python host code. This is entirely transparent
to the developer, who never interacts with \CUDAGDB or \PDB directly.

The session begins by setting breakpoints and issuing \CodeIn{run}
(line~\ref{exp:run}). Execution first halts at a \Python-side breakpoint
(line~\ref{exp:pyPolicyLineHit}), where the developer inspects \CodeIn{y}
(line~\ref{exp:PyPrintY}) and then continues. Execution halts inside the
kernel at the \CodeIn{temp} variable definition (Figure~\ref{fig:yax-example},
line~\ref{line:pyTemp2LOC}); \CodeIn{continue}
(line~\ref{exp:PyContinue}) moves deeper into device execution.

Once inside the kernel, the session illustrates two advanced
capabilities.
First, array slicing
(lines~\ref{exp:kernelPrintY02}--\ref{exp:kernelPrintYjj2}): rather than
transferring arrays into \CUDAGDB or \PDB directly, \ourTool accepts standard
\Python range syntax and fetches only the requested slice, saving memory
bandwidth. The values match those printed earlier on the \Python side on
line~\ref{exp:PyPrintY}, confirming the kernel is receiving the correct
input.\footnote{Digits differ slightly as the default \CUDAGDB precision is
      greater than \PDB precision.}
Second, live code evaluation (line~\ref{exp:kernelDeviceSum}): \ourTool
evaluates the expression \CodeIn{device\_sum} directly on the \GPU, computing
the sum of \CodeIn{y\_view} without moving data to the host; the result is
cross-checked by evaluating the equivalent expression on the \Python side (line~\ref{exp:pySumY}).

The session also demonstrates kernel-specific behavior. By default, \ourTool
debugs the first thread, as shown on line~\ref{exp:kernelPrintJ0}, where
printing \CodeIn{j}, the thread ID, returns 0. The developer then switches the
\CUDA context to block 0, thread 10 (line~\ref{exp:kernelCudaThread010}) and
prints \CodeIn{j} again (line~\ref{exp:kernelPrintJ10}); its value now reads
10, confirming that \CodeIn{j} reflects the active thread's context.

\section{Technique and Implementation}

We describe the design and implementation of~\ourTool.  We begin with an
overview of the architecture
in \S\ref{tech:arch-overview-section}
and specific features are detailed in
\S\ref{tech:device-functions-section}--\S\ref{tech:concurrent-launch}.

\subsection{Architecture}
\label{tech:arch-overview-section}

\usetikzlibrary{positioning, calc, fit, backgrounds, arrows.meta}

\definecolor{archUser}{HTML}{BDD7EE}
\definecolor{archPdb}{HTML}{A8D5BA}
\definecolor{archCtrl}{HTML}{F4B183}
\definecolor{archTgt}{HTML}{B8B8B8}

\begin{figure*}[t]
  \centering
  \resizebox{\textwidth}{!}{%
  \begin{tikzpicture}[
      font=\sffamily\footnotesize,
      proc/.style={
        rectangle, rounded corners=3pt,
        draw=#1!80!black, fill=#1!30,
        minimum width=1.6cm, minimum height=0.8cm,
        align=center, line width=0.7pt, inner sep=3pt,
      },
      proc/.default={archTgt},
      arr/.style={
        -{Latex[length=1.6mm, width=1.2mm]},
        line width=0.7pt, color=black!60,
      },
      stepnum/.style={
        circle, draw=black!75, fill=white,
        inner sep=1pt, minimum size=0.38cm,
        font=\sffamily\bfseries\tiny,
        line width=0.6pt,
      },
      lbl/.style={font=\sffamily\tiny, align=center, text=black!65},
      ptylbl/.style={
        font=\sffamily\tiny\bfseries, fill=white,
        inner sep=1.5pt, align=center,
      },
    ]

    \node[proc=archUser, minimum width=1.4cm, minimum height=0.8cm]
      (user) at (-2.8, 0)
      {{\tiny\textbf{User}}\\[-1pt]{\tiny CLI / REPL}};

    \node[proc=archPdb] (pdbp) at (0, 0) {
      {\tiny\textbf{\pdbp}}\\[-1pt]{\tiny Python dbg}
    };

    \node[proc=archCtrl] (ctrl) at (4.2, 0) {
      {\tiny\textbf{controller}}\\[-1pt]{\tiny per-platform}
    };

    \node[proc=archTgt, minimum width=2.2cm] (tgt) at (8.4, 0) {
      {\tiny\textbf{Target Debugger}}\\[-1pt]%
      {\tiny gdb / cuda-gdb/rocgdb}
    };

    \draw[arr] ([yshift= 0.20cm]user.east) -- ([yshift= 0.20cm]pdbp.west)
      node[pos=0.25, stepnum, name=s1] {1};
    \draw[arr] ([yshift=-0.20cm]pdbp.west) -- ([yshift=-0.20cm]user.east)
      node[pos=0.25, stepnum, name=s6] {6};

    \node[lbl, above=2pt of s1] {python -m pkdb [\textit{prog}]};
    \node[lbl, below=2pt of s6] {responses};

    \draw[arr] ([yshift= 0.20cm]pdbp.east) -- ([yshift= 0.20cm]ctrl.west)
      node[pos=0.15, stepnum] {2}
      node[midway, above=2pt, lbl] {env snapshot, line mapping};
    \draw[arr] ([yshift=-0.20cm]ctrl.west) -- ([yshift=-0.20cm]pdbp.east)
      node[midway, below=2pt, lbl] {user prompt, responses};

    \node[ptylbl] at ($(pdbp.east)!0.5!(ctrl.west)$) {PTY\,1};

    \draw[arr] ([yshift= 0.20cm]ctrl.east) -- ([yshift= 0.20cm]tgt.west)
      node[pos=0.15, stepnum] {3}
      node[midway, above=2pt, lbl] {attach, set launch breakpoint};
    \draw[arr] ([yshift=-0.20cm]tgt.west) -- ([yshift=-0.20cm]ctrl.east)
      node[pos=0.15, stepnum] {5}
      node[midway, below=2pt, lbl] {breakpoint stop, notification};

    \node[ptylbl] at ($(ctrl.east)!0.5!(tgt.west)$) {PTY\,2};

    \draw[arr, bend left=24] (tgt.south)
      to node[pos=0.25, stepnum] {4}
         node[below=3pt, pos=0.5, lbl] {\PyKokkos dispatch (kernel launch)}
      (pdbp.south);

  \end{tikzpicture}%
  }
  \caption{\ourTool{} architecture.  Three cooperating processes communicate
    over two pseudo-terminals (PTY).  The user launches \pdbp{}, our extension of
    the standard \Python debugger (Step~1).  On reaching a \PyKokkos parallel
    dispatch, \pdbp{} spawns the controller and establishes PTY\,1 (Step~2).  The
    controller spawns and attaches the target debugger over PTY\,2 (Step~3).
    \pdbp{} then resumes and \PyKokkos dispatches the kernel; the attached target
    debugger intercepts the launch (Step~4) and notifies the controller at
    breakpoints (Step~5).  Subsequent user commands are relayed from \pdbp{}
    through the controller to the target debugger; responses follow the same path
    in reverse (Step~6).}
  \label{fig:architecture}
\end{figure*}
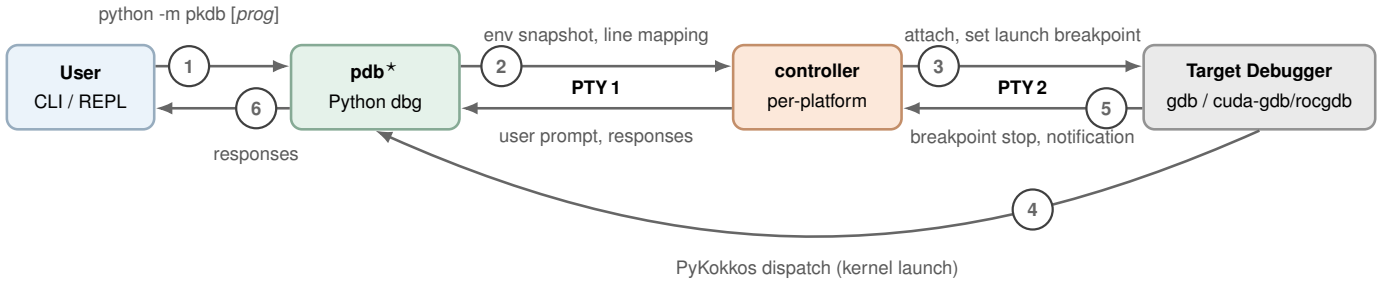

\ourTool is built around three cooperating processes: \pdbp, a controller, and a
target debugger.  \pdbp---our extension of the standard \Python debugger \PDB~\cite{python_pdb_module}---communicates with the controller over a Pseudo-Terminal
(PTY); the controller in turn communicates with the target debugger over a
second PTY\@.  Figure~\ref{fig:architecture} shows the resulting structure.  We
describe each step of a session below.

\MyPara{Step~1: \pdbp}
Execution begins when the user launches \ourTool from the CLI via
\CodeIn{python -m pkdb [program]},
which starts \pdbp;
additional commands are handled through
\PDB's well-established \CodeIn{cmd.Cmd} command interpreter protocol
\cite{python_cmd_module} (without modification)---the interface is immediately familiar to \Python developers.

\MyPara{Step~2: Entering a parallel dispatch}
Four preparatory actions are performed when execution reaches one of the \PyKokkos
parallel operations (\CodeIn{parallel\_for},
\CodeIn{parallel\_reduce}, \CodeIn{parallel\_scan}):

\begin{enumerate}[topsep=5pt,itemsep=2pt,partopsep=0ex,parsep=0ex,leftmargin=*]

      \item \textbf{Environment collection.}  \ourTool snapshots the \Python
            environment---global variable names, available
            functions, imported modules---at the call site and passes the snapshot to the
            controller so that \Python expressions can be evaluated directly from the device.

      \item \textbf{Execution-space discovery.}
            Arguments supplied to the parallel operation
            determine the effective execution space (e.g., \CodeIn{\CUDA},
            \CodeIn{HIP}, \CodeIn{OpenMP}) and \ourTool selects the
            appropriate controller;
            because this
            discovery is performed independently at each parallel dispatch, a
            single \ourTool session can debug a program that mixes
            execution spaces (e.g., a workload that dispatches some
            kernels to \OpenMP and others to \CUDA) without restarting the
            debugger.

      \item \textbf{Debug-mode compilation.}  The kernel is compiled by \PyKokkos with debug flags and optimizations disabled
            so that all debug information
            is preserved in the resulting binary; compiling without optimization is the standard approach during debugging.

      \item \textbf{Line-mapping construction.}  \ourTool builds a mapping from each
            \PyKokkos kernel line to the corresponding line in the produced intermediate
            C++, which was linked to compiled binary in the previous step;
            the mapping is used to translate user-specified breakpoints for the target debugger.

\end{enumerate}

After preparation, \pdbp spawns a controller process, passes it the collected
environment and line mapping, and establishes the first PTY channel for
communication.  The choice of controller implementation depends on the execution
space identified above: for example, an \CodeIn{\OpenMP} dispatch uses a
\CodeIn{GDBController}, while a \CodeIn{\CUDA} or \CodeIn{\HIP} dispatch uses an
\CodeIn{AcceleratorGDBController}.  The controller is the only component that must
be implemented per platform---all other parts of the architecture
are platform-agnostic---and each controller must implement the common controller API (Table~\ref{fig:controller-api}).
Table~\ref{tab:controllers} lists
implemented controllers
and corresponding debuggers, as well as the platforms they support.

\begin{table}[t]
      \centering
      \caption{Controller API shared across targets in \texttt{pkdb}\label{fig:controller-api}.}
      \Description{Table listing Controller methods shared across pkdb targets.}
      \footnotesize
      \begin{tabular}{@{}p{0.4\linewidth}p{0.55\linewidth}@{}}
            \toprule
            \textbf{API}                     & \textbf{Description}                        \\
            \midrule
            \texttt{start\_attached()}       & Start an attached debugger session         \\
            \texttt{stop()}                  & Terminate debugger and stop the controller \\
            \texttt{\_setup\_debugger()}     & Initialize debugger process/PTY wiring     \\
            \texttt{send\_command()}         & Send one command to the target debugger    \\
            \texttt{\_readline\_debugger()}  & Read one line of debugger output           \\
            \texttt{\_transparent\_loop()}   & Relay output/events while in debug mode    \\
            \texttt{\_command\_mode()}       & Run interactive user-input mode            \\
            \texttt{\_read\_command\_line()} & Read/parse a user command line             \\
            \texttt{parse\_and\_execute()}   & Parse and dispatch command                 \\
            \texttt{command()}               & Append function with command dispatch      \\
            \bottomrule
      \end{tabular}
\end{table}

\begin{table}[t]
  \centering
  \caption{Implemented \ourTool controllers with the associated target debugger and the supported \PyKokkos execution spaces (platforms).}
  \label{tab:controllers}
  \begin{tabular}{@{}lll@{}}
    \toprule
    \textbf{Controller} & \textbf{Target debugger} & \textbf{Platform} \\
    \midrule
    \texttt{GDBController} & \texttt{gdb} & \OpenMP{} \\
    \multirow{2}{*}{\texttt{AcceleratorGDBController}}
      & \texttt{cuda-gdb} & \CUDA{}\\
      & \texttt{rocgdb} & \HIP{} / ROCm \\
    \bottomrule
  \end{tabular}
\end{table}

\MyPara{Step~3: Spawning the target debugger}
The controller spawns  a platform-provided \emph{target debugger} (e.g., \CUDAGDB or \ROCGDB) and attaches it to the
\pdbp process.  A second PTY is established between the controller and
the target debugger for all subsequent communication.  The controller also
configures the target debugger to break execution at kernel launch, so that
control returns to \pdbp immediately when the kernel begins.

\MyPara{Step~4: Kernel launch}
\pdbp triggers the kernel launch through the unmodified \PyKokkos dispatch path.
As soon as the kernel starts, the
target debugger fires the launch breakpoint and notifies the controller, which
processes the notification, checks if there is a debugging point set
at the beginning of the kernel, and instructs the target debugger to continue to
the first breakpoint inside the kernel.

\MyPara{Step~5: Breakpoint stops}
The target debugger stops at a breakpoint and notifies the
controller, which consults the line mapping and the set of
user-defined breakpoints to determine whether the stop corresponds to a
user breakpoint or execution error (e.g., an illegal memory
access); in either case, the controller suspends execution and waits for user
input.

\MyPara{Step~6: User interaction}
The user issues commands through the \pdbp, which is redirected to the
controller; supported target commands are listed in Table~\ref{tab:commands}.
The controller interprets each command, translates it into the appropriate
instruction for the target debugger, and relays the response back to the user;
default pdb commands are sent directly to pdb.
Subsequent sections describe several of our custom commands.

\begin{table}[t]
      \centering
      \caption{%
            \ourTool{} commands exposed through \pdbp{} and handled by the
            controllers in Table~\ref{tab:controllers}. Commands marked with
            \textbullet~apply to both \OpenMP{} and \CUDA{}/\HIP{} sessions
            unless otherwise noted. Rows colored \colorbox{blue!15}{blue}
            denote commands that exist in both \PDB and \ourTool; rows
            colored \colorbox{green!22}{green} denote commands unique to
            \ourTool.%
      }
      \label{tab:commands}
      \begin{tabular}{@{}p{0.28\linewidth}p{0.49\linewidth}>{\centering\arraybackslash}m{0.1\linewidth}@{}}
            \toprule
            \textbf{Command}                         & \textbf{Role}                        & \textbf{Platform} \\
            \midrule
            \rowcolor{blue!15}
            \CodeIn{break}/\CodeIn{delete}           & Put breakpoint on workunit lines     & \textbullet       \\
            \rowcolor{blue!15}
            \CodeIn{continue}/\CodeIn{step}          & Continue execution                   & \textbullet       \\
            \rowcolor{blue!15}
            \CodeIn{quit}                            & Return to \pdbp                      & \textbullet       \\
            \rowcolor{blue!15}
            \CodeIn{eval}/\CodeIn{print}             & Evaluate expression / print variable & \textbullet       \\
            \rowcolor{green!22}
            \CodeIn{locals}                          & Show local variables                 & \textbullet       \\
            \rowcolor{green!22}
            \CodeIn{bt}/\CodeIn{whereami}            & Show backtrace                       & \textbullet       \\
            \rowcolor{green!22}
            \CodeIn{parallel\_print}                 & Print across all threads             & \textbullet       \\
            \rowcolor{green!22}
            \CodeIn{hotswap}                         & Change kernel invocation             & \textbullet       \\
            \rowcolor{green!22}
            \CodeIn{parallel\_launch}                & Parallel launch of multiple kernels  & \textbullet       \\
            \rowcolor{green!22}
            \CodeIn{args}                            & Show frame arguments                 & \textbullet       \\
            \rowcolor{green!22}
            \CodeIn{threads}/\CodeIn{thread}         & Print / select GDB thread(s)         & {OMP}             \\
            \rowcolor{green!22}
            \CodeIn{cuda}/\CodeIn{roc} \CodeIn{info} & Show framework-tied command(s)       & {CUDA ROCm}       \\
            \rowcolor{blue!15}
            \CodeIn{help}/\CodeIn{info}              & Show command help                    & \textbullet       \\
            \rowcolor{green!22}
            \CodeIn{send}                            & Send command directly to debugger    & \textbullet       \\
            \rowcolor{blue!15}
            \CodeIn{set}                             & Set debug properties                 & \textbullet       \\
            \rowcolor{blue!15}
            \CodeIn{len}                             & Print \CodeIn{View} size             & \textbullet       \\
            \bottomrule
      \end{tabular}
\end{table}

\subsection{\Liveeval}
\label{tech:device-functions-section}
\ourTool allows for the
evaluation of any arbitrary \Python expression at any point in a session; the
expression is executed as \Python code in the environment collected at the
parallel-dispatch boundary (Step~2 of \S\ref{tech:arch-overview-section}), so it
has access to the live global variables, functions, and modules at the
breakpoint: a non-array assignment made by the evaluated expression (e.g.,
rebinding a global name) is visible to the rest of the program once execution
resumes, just as array mutations are (the array case, described below, requires
additional handling because array data may live outside \pdbp's own memory). The
syntax for \CodeIn{eval} can be found in Figure~\ref{fig:debug-experience}
line~\ref{exp:kernelDeviceSum}: the user writes \CodeIn{eval} followed by the
evaluated command,
where \CodeIn{f} is any callable in the captured environment and \CodeIn{arr} is
a \cupy array. \CodeIn{eval} may call an ordinary \Python function, or it may
dispatch a new \PyKokkos kernel, hence \ourTool effectively supports live kernel
invocation at a stop without any modification of the debugged program.

When an expression argument is an array,
\ourTool
passes it to the evaluation context without copying the underlying
data---for a device array, copying large device buffers to the host is slow, and, for
datasets that already fill device memory, can trigger out-of-memory
errors; avoiding the copy eliminates both problems.

\ourTool avoids device-to-host copies via
\GPU IPC memory handles:
for each array argument, \pdbp obtains an IPC HandleInfo by calling
\CodeIn{ipcGetMemHandle} on the array's device pointer and transmits the
HandleInfo to the controller; the controller opens the HandleInfo with
\CodeIn{ipcOpenMemHandle}, constructs a \cupy array view over the same device
memory, and uses that view when invoking the expression. Because both processes
alias the same device allocation, writes made by the evaluated expression are
visible to the rest of the program; \ourTool does not restrict mutability, that
choice is left to the user. For host arrays on the \OpenMP execution space, the
shared memory region is accessed directly (i.e., without an IPC HandleInfo).

\begin{algorithm}[t]
      \caption{Invocation of device side during live code evaluation}
      \label{algo:deviceFunctionsAlgo}
      \begin{algorithmic}[1]
            \State \textbf{Input:} Device function name $f_{\text{name}}$ (e.g., ``device\_sum''), arguments $\mathcal{D}_{\text{args}}$.
            \State \textbf{Output:} Result of device-side execution.

            \vspace{0.2em}

            \State $\mathcal{D}_{\text{dev}} \leftarrow [\ ]$, $\;\mathcal{P}_{\text{dev}} \leftarrow \emptyset$
            
            \For{each argument $d \in \mathcal{D}_{\text{args}}$}\label{dse:for}
            \If{ $\neg$\CodeIn{hasattr(d, data)} or $\neg$\CodeIn{hasattr(d.data, ptr)} }
            \State $\mathcal{D}_{\text{dev}}\text{.append}(d)$
            \State continue
            \EndIf

            \State $\text{ptr} \leftarrow d.\text{data.ptr}$
            \State $h_{\text{IPC}} \leftarrow \text{cupy.cuda.runtime.ipcGetMemHandle}(\text{ptr})$

            \State $\text{ptr}_{\text{dev}} \leftarrow \text{ipcOpenMemHandle}(h_{\text{IPC}}.hex)$ \label{dse:open}
            \State $d_{\text{dev}} \leftarrow \text{cupy.ndarray}(d.\text{shape} , d.\text{dtype}, \text{ptr}_{\text{dev}})$ \label{dse:cupy}
            \State $\mathcal{D}_{\text{dev}}\text{.append}(d_{\text{dev}})$
            \State $\mathcal{P}_{\text{dev}} \leftarrow \mathcal{P}_{\text{dev}} \cup \{\text{ptr}_{\text{dev}}\}$
            \EndFor\label{dse:endfor2}
            \State $\text{result} \leftarrow \text{pykokkos.parallel\_\textless op\textgreater}(f_{\text{name}}, \mathcal{D}_{\text{dev}})$ \label{dse:fun}
            \For{each $\text{ptr}_{\text{dev}} \in \mathcal{P}_{\text{dev}}$}
            \State $\text{ipcCloseMemHandle}(\text{ptr}_{\text{dev}})$
            \EndFor
            \State \Return $\text{result}$

      \end{algorithmic}
\end{algorithm}

Algorithm~\ref{algo:deviceFunctionsAlgo} formalizes the IPC-based
array-passing mechanism.  \ourTool iterates over the call's arguments
(lines~\ref{dse:for}--\ref{dse:endfor2}), skipping any argument that is
not a device array; for each device array, it retrieves the device
pointer, requests an IPC memory handle for it, and immediately opens
that handle (line~\ref{dse:open}), reconstructing a \cupy array view
over the same device allocation (line~\ref{dse:cupy}). The resulting
view is added to the argument list $\mathcal{D}_{\text{dev}}$ and the
opened device pointer to $\mathcal{P}_{\text{dev}}$ set. Once every argument
has been processed, \ourTool dispatches the function once with the
collected set of \cupy views (line~\ref{dse:fun}) and, once the call
returns, closes all opened IPC handles before returning the result.

\subsection{\KernelSubstitute}
\label{tech:kernel-hotswapping-section}
\begin{algorithm}[t]
      \caption{\KernelSubstitute}
      \label{algo:kernelHotswapAlgo}
      \begin{algorithmic}[1]

            \State Global map \substGlobalMap and per-line map
            \substLocalMap store substitutions \substMapping. \label{khs:map}

            \Statex \hrulefill
            \State \textbf{On}~\CodeIn{hotswap} \substituteFrom \substituteTo [$\substLineFunc$]: store \substMapping
            in \substGlobalMap or in \substLocalMap for \substLine. \label{khs:hotswap}
            \Statex \hrulefill
            \State \textbf{On} dispatch of a $\parallelOpsCode$ call at line
            \substLine: let \substituteFrom be the invoked function; if \substituteFrom~$\in$~\substGlobalMap
            or (\substituteFrom, \substLine)~$\in$~\substLocalMap, call
            \Call{Apply}{\substituteFrom,\substLine}. \label{khs:dispatch}
            \Statex \hrulefill

            \Function{Apply}{\substituteFrom,\substLine}\label{khs:apply}
            \State $\CodeLine \leftarrow$ source text at \substLine;
            \State \substNewState~$\leftarrow$~\substGlobalMap updated by
            \substLocalMapEntry if defined. \label{khs:merge}
            \State \substituteTo~$\leftarrow$~\newStateGetS. \label{khs:check}

            \If {signature(\substituteTo) $\ne$ signature(\substituteFrom)} \label{khs:sigMatchStart}
                  \State \Return \argMismatch
            \EndIf \label{khs:sigMatchEnd}
            
            \State Build $\parallelOpsCode'$ with \substituteTo instead of \substituteFrom in \CodeLine; \label{khs:ex}
            \State run $\parallelOpsCode'$ and get result \substResult;
            \State \Return \substResult
            \EndFunction

      \end{algorithmic}
\end{algorithm}

\KernelSubstitute lets the user replace kernels
during a debug session without restarting the process.  This
is especially valuable in long-running \HPC jobs, where a restart
means reloading inputs, reallocating device memory, warming caches,
and losing the failure point under inspection.

Some CPU debuggers, such as \GDB, support reverse execution and time-travel
debugging, which can be combined with on-the-fly
patching~\cite{boric_reverse-engineering_2023}. However, accelerator-oriented
debuggers such as \CUDAGDB expose a narrower feature set and do not support
record/replay for reverse debugging or similar mechanisms.
Instead of using record/replay, \ourTool relies on the dynamic nature of
\PyKokkos, where kernels are compiled and loaded on demand at runtime---a
capability that, to our knowledge, no existing technique has exploited---and uses
call-site interposition~\cite{vmmCallInterposition,onFlightPatching}: at each
\parallelOpsCode dispatch, \ourTool checks whether the callee should be replaced
and, if so, redirects the call without touching the \Python source file on disk
by changing the \Python's reference to called function.

\MyPara{Commands}
\ourTool provides two forms of the \CodeIn{hotswap} command for \kernelSubstitute:
\begin{enumerate*}
      \item \CodeIn{hotswap kernel1 kernel2} {
                  \noindent registers a global substitution---every future dispatch that
                  would invoke \CodeIn{kernel1} is redirected to \CodeIn{kernel2}
                  instead;}
      \item \CodeIn{hotswap lineno kernel1 kernel2} {
                  \noindent restricts the substitution to the single call site at
                  \CodeIn{lineno}, leaving all other dispatches of \CodeIn{kernel1}
                  unchanged.
            }
\end{enumerate*}

The replacement kernel in both commands must match the signature of the
original; if the signatures differ, \ourTool reports an error and falls back to
the original kernel.

\MyPara{Mechanism}
Algorithm~\ref{algo:kernelHotswapAlgo} formalizes how substitutions are recorded
and applied.  Global substitution map \substGlobalMap stores entries. Each entry
includes function (\substituteFrom), that should be substituted and function
(\substituteTo), that we should substitute to; The map \substLocalMap
(line~\ref{khs:map}) additionally stores the location \substLine at which
\substituteFrom should be substituted by \substituteTo. \Call{Hotswap}{} records
the substitution. Dispatch of a $\parallelOpsCode$ (line~\ref{khs:hotswap}) call
checks whether the invoked function~\substituteFrom has a registered
substitution in either map and, if so, invokes \Call{Apply}{}
(line~\ref{khs:dispatch}). Because \Call{Hotswap}{} never modifies the source
file, a call site keeps invoking the same original function \substituteFrom on
every subsequent dispatch, so \Call{Apply}{} (line~\ref{khs:apply}) must consult
the maps again at each dispatch, building global state $\mathcal{H}$, which
represents all currently-substituted functions, to resolve what \substituteFrom
should be replaced with right now. It is done by merging functions from
\substGlobalMap overridden by any line-specific entry from \substLocalMap for
(\substituteFrom, \substLine) (line~\ref{khs:merge}), so a per-line substitution
always takes precedence over a global one at that call site; it then looks up
the replacement \substituteTo $\leftarrow$~\newStateGetS (line~\ref{khs:check}),
checks for signatures match
(lines~\ref{khs:sigMatchStart}--\ref{khs:sigMatchEnd}), and executes
\substituteTo in place of \substituteFrom (line~\ref{khs:ex}).

\subsection{Concurrent kernel launch}
\label{tech:concurrent-launch}

At a breakpoint, a user may wish to run several kernels simultaneously rather
than one at a time.  Running kernels concurrently saves time and, importantly,
allows their results to be compared directly---for example, to check whether two
implementations of the same computation agree on the live data at a
breakpoint. Additionally, kernels in different clauses may target different
execution spaces, so that, e.g., variants running on \GPU and \CPU can be
compared in a single command.

The command syntax consists of multiple clauses:
\CodeIn{parallel\_launch~$c_{1}$;~$c_{2}$;~...; $c_{n}$}, where each $c_{i}$
clause specifies a parallel operation, a kernel name, an optional execution space
(with the \CodeIn{pk.} prefix, e.g., \CodeIn{pk.Cuda}),
\CodeIn{RangePolicy}, and arguments.  All names and expressions are resolved via
\Python \CodeIn{eval} against the live environment collected at the
parallel-dispatch boundary, so symbols such as \CodeIn{N} or array objects refer
to current program state.  For example, the \CodeIn{yAx} kernel from
Figure~\ref{fig:yax-python} can be launched on \GPU and on \OpenMP with
\CodeIn{parallel\_launch yAx pk.Cuda pk.RangePolicy(0,N) y x A; yAx pk.OpenMP pk.RangePolicy(0,N) y x A};
\noindent a failure in one clause is reported but does not prevent the
remaining clauses from executing.

By default, array arguments are treated as read-only: each clause
receives an isolated copy of the data so that concurrent kernels do
not interfere.  For \cupy arrays (\GPU execution spaces), the copy is
taken as a host snapshot; for \NumPy arrays (\OpenMP execution
spaces), a separate buffer is allocated on the \CPU\@.

A user can opt out of this isolation by marking an argument as \emph{mutable}
using the \CodeIn{mut:} prefix (e.g., \CodeIn{mut:arr} or
\CodeIn{mut:arr=<expr>}).  Mutable \cupy arrays are shared across processes via
IPC without copying, exactly as in live code evaluation
(\S\ref{tech:device-functions-section}); mutable \NumPy arrays are passed by
reference so the kernel and the parent alias the same storage.  If the same
buffer is marked mutable in multiple concurrent clauses, races and silent
corruption are possible; \ourTool applies the annotations directly and leaves
correctness to the user.

\subsection{Thread-aware continue}
\label{tech:thread-stepping-section}

By default, when debugging \OpenMP kernels with \GDB, a \CodeIn{continue}
command resumes all threads concurrently. As a result, a thread can reach
breakpoint faster than others and will forcefully stop other threads, even if
they are not at the breakpoint positions. Algorithm~\ref{algo:continue-native}
specifies how \ourTool mitigates this issue, by locking the \GDB scheduler
(line~\ref{npec:lock}) and continuing threads in $\mathcal{T}$ explicitly
(line~\ref{npec:cont}), one by one, rather than letting them race, so the
resulting state is deterministic; once all threads reaches a breakpoint,
\ourTool releases the lock, restoring \GDB's default (unlocked) scheduling
behavior.

Another limitation of \GDB is that the default \CodeIn{continue} always resumes
every thread; the user cannot resume an arbitrary subset. To give the user this
control, \ourTool provides an extended continue command.
\CodeIn{continue~threads~[begin:end]},
\noindent which resumes only the specified range. Issuing \CodeIn{continue}
without a range resumes all threads. \ourTool does not specifically handle
cases, like breakpoints inside of thread-dependent \CodeIn{if} branches.

\begin{algorithm}[t]
      \caption{Thread-aware continue command}
      \label{algo:continue-native}
      \begin{algorithmic}[1]
            \State \textbf{Input:} Breakpoint set $\mathcal{B}$; active threads $\mathcal{T}$.
            \State \CodeIn{gdb-set scheduler-locking on} \label{npec:lock}
            \State send \CodeIn{continue} to all threads in $\mathcal{T}$ \label{npec:cont}
            \State Wait until all $\mathcal{T}$ reaches a location in $\mathcal{B}$
            \State \CodeIn{gdb-set scheduler-locking off} \label{npec:lock-on}
      \end{algorithmic}
\end{algorithm}

\section{Evaluation}
\label{sec:evaluation}

We evaluate the user-independent factors of \ourTool by performing experiments to answer the following questions:

\begin{enumerate}[leftmargin=*, itemsep=0.25ex, topsep=0.25ex, parsep=0pt, partopsep=0pt]
      \item { {\bf Debug overhead}
            (\S\ref{eval:benchmarks-evaluation-section}): How does wall time
            differ between \ourTool's debug mode versus \PDB debug execution, on
            \CodeIn{\CUDA}, \CodeIn{\HIP} and \CodeIn{\OpenMP} backends?  }
      \item {
            {\bf Comparison with the \PyKokkos debug mode}
            (\S\ref{eval:debug-time-comparison-section}):
            The existing \PyKokkos Debug execution mode lowers all parallel
            work to serial, Python-only execution for
            debugging; how do these times differ
            compared to \ourTool, which dispatches to the unmodified
            \GPU/\OpenMP backend during debugging?
            }
      \item {
            {\bf \KernelSubstitute cost}
            (\S\ref{eval:hot-swapping-section}):
            What is the wall-clock cost of substituting a kernel call by editing the source and
            restarting the session (\EditDebug) versus issuing \CodeIn{hotswap}
            inside a running session ({call site substitution})?
            }
\end{enumerate}

\noindent
Additionally, we provide two case studies (\S\ref{eval:case-studies-section}) where \ourTool is used to debug two \PyKokkos research applications: a Boltzmann particle-in-cell kinetics code and an Ewald summation code.

We describe the evaluation setup (\S\ref{eval:setup}) and subjects
(\S\ref{eval:subjects}) before addressing each experiment
(\S\ref{eval:benchmarks-evaluation-section}-\S\ref{eval:hot-swapping-section}) and the case studies (\S\ref{eval:case-studies-section}).

\subsection{Evaluation setup}
\label{eval:setup}

\begin{table}[t]
  \centering
  \footnotesize
  \caption{Machines used in evaluation. For \CPU evaluation, we set
    \CodeIn{OMP\_NUM\_THREADS} to the number of available \CPU cores.}
  \label{tab:machines}
  \rowcolors{1}{white}{gray!15}
  \begin{tabularx}{\columnwidth}{l X X}
    \toprule
    \textbf{Machine} & \textbf{\CPU}                 & \textbf{\GPU}                      \\
    \midrule
    \Tokyo           & \Intel Xeon~w5, 32 cores      & \NVIDIA RTX 5000, 32 GB            \\
    \Vista           & \NVIDIA Grace, 72 cores       & \NVIDIA H200, 96 GB                \\
    \amdCluster      & \AMD EPYC, 96 cores & \AMD MI300X, 192 GB      \\
    \bottomrule
  \end{tabularx}
\end{table}

Experiments are run on the machines listed in Table~\ref{tab:machines}.
Each experiment is run four times, with a dry run discarded and results reported
as averages over the remaining times. Moreover, for
Section~\ref{eval:benchmarks-evaluation-section} we disabled \ourTool's
optimization that skips kernel debugging if there is no breakpoint inside of the
kernel. This lets us measure the overhead of the full \ourTool framework, not
just \pdbp{} in isolation.

\subsection{Subjects}
\label{eval:subjects}

We briefly describe subjects used in our evaluation.

\MyPara{\Exa}
The \Exa mini-application
is a modular codebase designed to investigate
performance of common kernels in particle code
frameworks~\cite{examinimd_2025}.
\Exa was ported from \Kokkos C++ to \PyKokkos~\cite{PyKokkos}, where
the total execution time was shown to be competitive.
Given an $N$ atom system, the execution flow
comprises the following steps:
\begin{enumerate*}
      \item \CodeIn{Initialize} position, velocity, and force arrays of size
            $3N$;
      \item \CodeIn{Compute} temperature, potential energy, and kinetic
            energy of the system using \PyKokkos parallel kernels;
      \item \CodeIn{Update} position, velocity, and force arrays;
      \item \CodeIn{Repeat} steps 2 and 3 for 100 time steps.
\end{enumerate*}

\MyPara{\Boltz}
The \Boltz kinetic equations solver is a particle-in-cell
code that uses \NumPy/\CuPy for linear algebra and \PyKokkos
for parallel kernels
\cite{Almgren-Bell_Awar_Geethakrishnan_Gligoric_Biros_2022}.  The
scheme has three main steps:
\begin{enumerate*}
      \item \CodeIn{Collision}: compute collision probabilities, requires on-device random number
            generation, reshuffling in memory and atomics;
      \item \CodeIn{Recombination}: compute a three-body problem, requires two
            distinct kernels with low arithmetic intensity and atomics;
      \item \CodeIn{Particle-to-Cell}: compute per-cell
            right-hand-side vectors, requires a high arithmetic intensity
            kernel with atomics.
\end{enumerate*}

\MyPara{\Ewald}
Ewald summation---used to accelerate molecular dynamics simulations and potential theory calculations---comprises the following steps:
\begin{enumerate*}
      \item \CodeIn{P2P}: interactions between
            particles within a cutoff;
      \item \CodeIn{P2G}: spread particle forces onto the grid by convolving
            with a compactly supported window function;
      \item \CodeIn{FFT}: compute Fourier space grid forces;
      \item \CodeIn{CNV}: apply convolution in Fourier space
            for the grid velocities;
      \item \CodeIn{IFFT}: compute real space grid velocities;
      \item \CodeIn{G2P}: interpolate the grid velocities to
            particles.
\end{enumerate*}
The algorithmic implementation, given in \cite{parki}, uses \PyKokkos for
parallel kernels and the \NumPy/\CuPy libraries for data structures and FFTs.
The \CodeIn{P2P}, \CodeIn{P2G} and \CodeIn{CNV} provide representative \HPC
benchmarks: \CodeIn{P2P} has a high arithmetic intensity and a spatially local
cache structure, \CodeIn{P2G} uses shared memory and atomic operations, and
\CodeIn{CNV} has a low arithmetic intensity.

\subsection{Debug overhead}
\label{eval:benchmarks-evaluation-section}

\begin{figure*}[t]
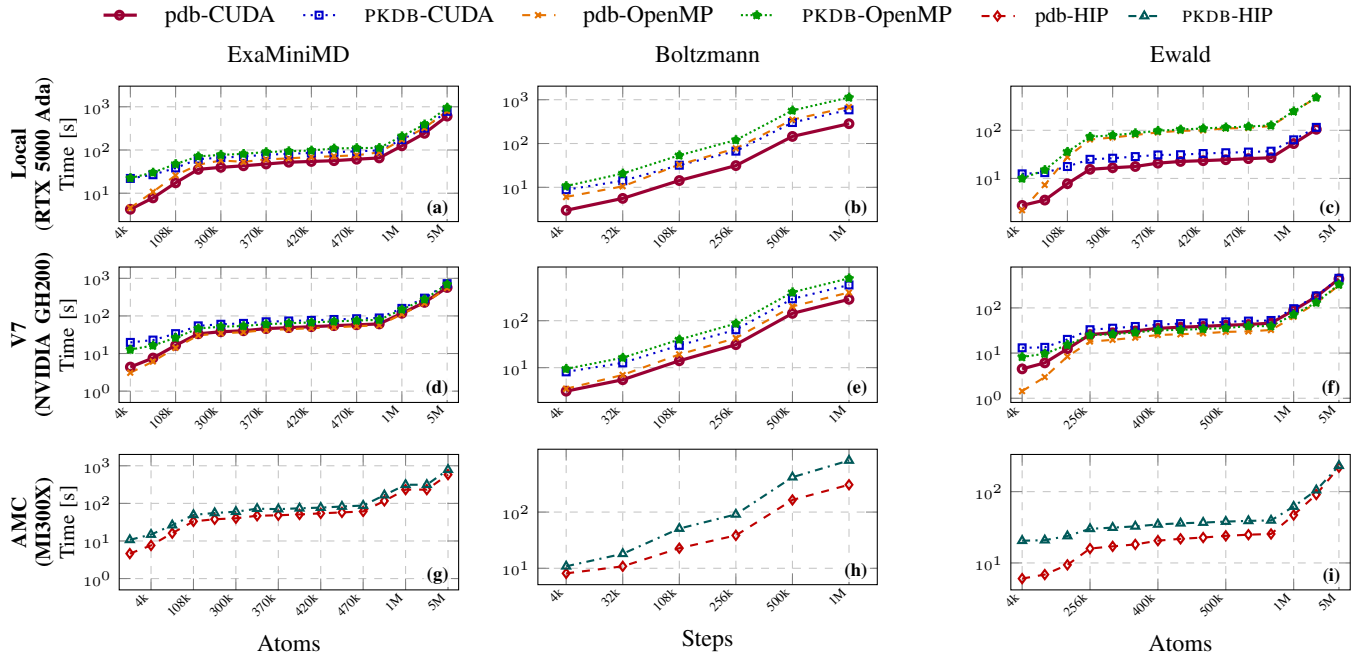

  \centering
  \footnotesize
  \captionsetup[subfigure]{skip=2pt}
  \newcommand{\EvalCombH}{0.6\linewidth}
  \newcommand{\EvalCombW}{1.08\linewidth}

  \ref{eval-legend-a}~\ref{eval-legend-amd}\par\vspace{0.35ex}
  \begin{subfigure}[t]{0.31\textwidth}
    \centering
    \begin{tikzpicture}
      \begin{axis}[
          width=\EvalCombW,
          height=\EvalCombH,
          xlabel={},
          title={ExaMiniMD},
          title style={font=\small},
          ylabel={\textbf{\Tokyo{}}\\(\textbf{RTX 5000 Ada})\\Time [s]},
          xmin=0.5,
          xmax=15.5,
          xtick={{ 1, 3, 5, 7, 9, 11, 13, 15, 17 }},
          xticklabels={{ 4k, 108k, 300k, 370k, 420k, 470k, 1M, 5M, 10M }},
          xticklabel style={rotate=45, anchor=east, font=\tiny},
          yticklabel style={font=\tiny},
          ylabel style={font=\footnotesize, align=center, at={(axis description cs:-0.22,0.5)}, anchor=center},
          xmajorgrids=true,
          ymajorgrids=true,
          grid style={dashed,gray!\chartDashedLineOpacity},
          scaled ticks=false,
          legend cell align={left},
          legend to name=eval-legend-c,
          legend style={legend columns=-1, font=\small, draw=none, column sep=1em},
          ymode=log,
          log basis y={10},
          ymax=3000,
          ytickten={-1,0,1,2,3},
          minor y tick num=0,
        ]
        \input{figures/eval-ExaMiniMD-exectime-data.tex}
        \node[anchor=south east, font=\scriptsize\bfseries, fill=white, fill opacity=0.85, text opacity=1, inner sep=1pt]
          at (rel axis cs:0.98,0.02) {(a)};
      \end{axis}
    \end{tikzpicture}
    \phantomsubcaption
    \label{fig:eval-examinimd-exectime}
  \end{subfigure}%
  \hfill
  \begin{subfigure}[t]{0.31\textwidth}
    \centering
    \begin{tikzpicture}
      \begin{axis}[
          width=\EvalCombW,
          height=\EvalCombH,
          xlabel={},
          title={Boltzmann},
          title style={font=\small},
          ylabel={},
          xmin=0.5,
          xmax=6.5,
          xtick={{ 1, 2, 3, 4, 5, 6 }},
          xticklabels={{ 4k, 32k, 108k, 256k, 500k, 1M }},
          xticklabel style={rotate=45, anchor=east, font=\tiny},
          yticklabel style={font=\tiny},
          legend cell align={left},
          legend to name=eval-legend-b,
          legend style={legend columns=-1, font=\small, draw=none, column sep=1em},
          ymode=log,
          log basis y={10},
          ytickten={-1,0,1,2,3},
          minor y tick num=0,
          xmajorgrids=true,
          ymajorgrids=true,
          grid style={dashed,gray!\chartDashedLineOpacity},
          scaled ticks=false,
        ]
        \input{figures/eval-boltzmann-data.tex}
        \node[anchor=south east, font=\scriptsize\bfseries, fill=white, fill opacity=0.85, text opacity=1, inner sep=1pt]
          at (rel axis cs:0.98,0.02) {(b)};
      \end{axis}
    \end{tikzpicture}
    \phantomsubcaption
    \label{fig:boltzmann-exectime}
  \end{subfigure}%
  \hfill
  \begin{subfigure}[t]{0.31\textwidth}
    \centering
    \begin{tikzpicture}
      \begin{axis}[
          width=\EvalCombW,
          height=\EvalCombH,
          xlabel={},
          title={Ewald},
          title style={font=\small},
          ylabel={},
          xmin=0.5,
          xmax=15.5,
          xtick={{ 1, 3, 5, 7, 9, 11, 13, 15}},
          xticklabels={{ 4k, 108k, 300k, 370k, 420k, 470k, 1M, 5M }},
          xticklabel style={rotate=45, anchor=east, font=\tiny},
          yticklabel style={font=\tiny},
          legend cell align={left},
          legend to name=eval-legend-a,
          legend style={legend columns=-1, font=\small, draw=none, column sep=1em},
          ymode=log,
          log basis y={10},
          ytickten={-1,0,1,2,3},
          minor y tick num=0,
          xmajorgrids=true,
          ymajorgrids=true,
          grid style={dashed,gray!\chartDashedLineOpacity},
          scaled ticks=false,
        ]
        \input{figures/eval-ewald-data.tex}
        \node[anchor=south east, font=\scriptsize\bfseries, fill=white, fill opacity=0.85, text opacity=1, inner sep=1pt]
          at (rel axis cs:0.98,0.02) {(c)};
      \end{axis}
    \end{tikzpicture}
    \phantomsubcaption
    \label{fig:ewald-exectime}
  \end{subfigure}

  \vspace{-0.5cm}
  \begin{subfigure}[t]{0.31\textwidth}
    \centering
    \begin{tikzpicture}
      \begin{axis}[
          width=\EvalCombW,
          height=\EvalCombH,
          xlabel={},
          ylabel={\textbf{\Vista{}} \\ \textbf{(NVIDIA GH200)}\\Time [s]},
          xmin=0.5,
          xmax=15.5,
          xtick={{ 1, 3, 5, 7, 9, 11, 13, 15, 17 }},
          xticklabels={{ 4k, 108k, 300k, 370k, 420k, 470k, 1M, 5M, 10M }},
          xticklabel style={rotate=45, anchor=east, font=\tiny},
          yticklabel style={font=\tiny},
          ylabel style={font=\footnotesize, align=center, at={(axis description cs:-0.22,0.5)}, anchor=center},
          legend cell align={left},
          legend to name=eval-legend-gh200,
          legend style={legend columns=-1, font=\footnotesize, draw=none, column sep=0.65em},
          ymode=log,
          log basis y={10},
          ymin=0.5,
          ymax=2000,
          ytickten={-1,0,1,2,3},
          minor y tick num=0,
          xmajorgrids=true,
          ymajorgrids=true,
          grid style={dashed,gray!\chartDashedLineOpacity},
          scaled ticks=false,
        ]
        \input{figures/eval-ExaMiniMD-exectime-data-vista-gh200.tex}
        \node[anchor=south east, font=\scriptsize\bfseries, fill=white, fill opacity=0.85, text opacity=1, inner sep=1pt]
          at (rel axis cs:0.98,0.02) {(d)};
      \end{axis}
    \end{tikzpicture}
    \phantomsubcaption
    \label{fig:eval-examinimd-exectime-gh200}
  \end{subfigure}%
  \hfill
  \begin{subfigure}[t]{0.31\textwidth}
    \centering
    \begin{tikzpicture}
      \begin{axis}[
          width=\EvalCombW,
          height=\EvalCombH,
          xlabel={},
          ylabel={},
          xmin=0.5,
          xmax=6.5,
          xtick={{ 1, 2, 3, 4, 5, 6 }},
          xticklabels={{ 4k, 32k, 108k, 256k, 500k, 1M }},
          xticklabel style={rotate=45, anchor=east, font=\tiny},
          yticklabel style={font=\tiny},
          legend cell align={left},
          legend to name=eval-legend-gh200,
          legend style={legend columns=-1, font=\footnotesize, draw=none, column sep=0.65em},
          ymode=log,
          log basis y={10},
          ytickten={-1,0,1,2},
          minor y tick num=0,
          xmajorgrids=true,
          ymajorgrids=true,
          grid style={dashed,gray!\chartDashedLineOpacity},
          scaled ticks=false,
        ]
        \input{figures/eval-boltzmann-data-vista-gh200.tex}
        \node[anchor=south east, font=\scriptsize\bfseries, fill=white, fill opacity=0.85, text opacity=1, inner sep=1pt]
          at (rel axis cs:0.98,0.02) {(e)};
      \end{axis}
    \end{tikzpicture}
    \phantomsubcaption
    \label{fig:boltzmann-exectime-gh200}
  \end{subfigure}%
  \hfill
  \begin{subfigure}[t]{0.31\textwidth}
    \centering
    \begin{tikzpicture}
      \begin{axis}[
          width=\EvalCombW,
          height=\EvalCombH,
          xlabel={},
          ylabel={},
          xmin=0.5,
          xmax=15.5,
          xtick={{ 1, 4, 7, 10, 13, 15 }},
          xticklabels={{ 4k, 256k, 400k, 500k, 1M, 5M }},
          xticklabel style={rotate=45, anchor=east, font=\tiny},
          yticklabel style={font=\tiny},
          legend cell align={left},
          legend to name=eval-legend-gh200,
          legend style={legend columns=-1, font=\footnotesize, draw=none, column sep=0.65em},
          ymode=log,
          log basis y={10},
          ytickten={-2,-1,0,1,2,3},
          minor y tick num=0,
          xmajorgrids=true,
          ymajorgrids=true,
          grid style={dashed,gray!\chartDashedLineOpacity},
          scaled ticks=false,
        ]
        \input{figures/eval-ewald-data-vista-gh200.tex}
        \node[anchor=south east, font=\scriptsize\bfseries, fill=white, fill opacity=0.85, text opacity=1, inner sep=1pt]
          at (rel axis cs:0.98,0.02) {(f)};
      \end{axis}
    \end{tikzpicture}
    \phantomsubcaption
    \label{fig:ewald-exectime-gh200}
  \end{subfigure}

  \vspace{-0.2cm}
  \begin{subfigure}[t]{0.31\textwidth}
    \centering
    \begin{tikzpicture}
      \begin{axis}[
          width=\EvalCombW,
          height=\EvalCombH,
          xlabel={Atoms},
          ylabel={\textbf{\amdCluster{}} \\ \textbf{(MI300X)}\\Time [s]},
          xmin=0.5,
          xmax=16.5,
          xtick={{ 2, 4, 6, 8, 10, 12, 14, 16, 18 }},
          xticklabels={{ 4k, 108k, 300k, 370k, 420k, 470k, 1M, 5M, 10M }},
          xticklabel style={rotate=45, anchor=east, font=\tiny},
          yticklabel style={font=\tiny},
          xlabel style={font=\small},
          ylabel style={font=\footnotesize, align=center, at={(axis description cs:-0.22,0.5)}, anchor=center},
          legend cell align={left},
          legend to name=eval-legend-amd,
          legend style={legend columns=-1, font=\footnotesize, draw=none, column sep=0.65em},
          ymode=log,
          log basis y={10},
          ymin=0.5,
          ymax=2000,
          ytickten={-1,0,1,2,3},
          minor y tick num=0,
          xmajorgrids=true,
          ymajorgrids=true,
          grid style={dashed,gray!\chartDashedLineOpacity},
          scaled ticks=false,
        ]
        \input{figures/eval-ExaMiniMD-exectime-data-amd-mi300x.tex}
        \node[anchor=south east, font=\scriptsize\bfseries, fill=white, fill opacity=0.85, text opacity=1, inner sep=1pt]
          at (rel axis cs:0.98,0.02) {(g)};
      \end{axis}
    \end{tikzpicture}
    \phantomsubcaption
    \label{fig:eval-examinimd-exectime-amd}
  \end{subfigure}%
  \hfill
  \begin{subfigure}[t]{0.31\textwidth}
    \centering
    \begin{tikzpicture}
      \begin{axis}[
          width=\EvalCombW,
          height=\EvalCombH,
          xlabel={Steps},
          ylabel={},
          xmin=0.5,
          xmax=6.5,
          xtick={{ 1, 2, 3, 4, 5, 6 }},
          xticklabels={{ 4k, 32k, 108k, 256k, 500k, 1M }},
          xticklabel style={rotate=45, anchor=east, font=\tiny},
          yticklabel style={font=\tiny},
          xlabel style={font=\small},
          legend cell align={left},
          legend to name=eval-legend-amdr ,
          legend style={legend columns=-1, font=\small, draw=none, column sep=1em},
          ymode=log,
          log basis y={10},
          ytickten={-1,0,1,2},
          minor y tick num=0,
          xmajorgrids=true,
          ymajorgrids=true,
          grid style={dashed,gray!\chartDashedLineOpacity},
          scaled ticks=false,
        ]
        \input{figures/eval-boltzmann-data-amd-mi300x.tex}
        \node[anchor=south east, font=\scriptsize\bfseries, fill=white, fill opacity=0.85, text opacity=1, inner sep=1pt]
          at (rel axis cs:0.98,0.02) {(h)};
      \end{axis}
    \end{tikzpicture}
    \phantomsubcaption
    \label{fig:boltzmann-exectime-amd}
  \end{subfigure}%
  \hfill
  \begin{subfigure}[t]{0.31\textwidth}
    \centering
    \begin{tikzpicture}
      \begin{axis}[
          width=\EvalCombW,
          height=\EvalCombH,
          xlabel={Atoms},
          ylabel={},
          xmin=0.5,
          xmax=15.5,
          xtick={{ 1, 4, 7, 10, 13, 15 }},
          xticklabels={{ 4k, 256k, 400k, 500k, 1M, 5M }},
          xticklabel style={rotate=45, anchor=east, font=\tiny},
          yticklabel style={font=\tiny},
          xlabel style={font=\small},
          legend cell align={left},
          legend to name=eval-legend-amd,
          legend style={legend columns=-1, font=\footnotesize, draw=none, column sep=0.65em},
          ymode=log,
          log basis y={10},
          ytickten={-2,-1,0,1,2},
          minor y tick num=0,
          xmajorgrids=true,
          ymajorgrids=true,
          grid style={dashed,gray!\chartDashedLineOpacity},
          scaled ticks=false,
        ]
        \input{figures/eval-ewald-data-amd-mi300x.tex}
        \node[anchor=south east, font=\scriptsize\bfseries, fill=white, fill opacity=0.85, text opacity=1, inner sep=1pt]
          at (rel axis cs:0.98,0.02) {(i)};
      \end{axis}
    \end{tikzpicture}
    \phantomsubcaption
    \label{fig:ewald-exectime-amd}
  \end{subfigure}

  \caption{%
    Wall time for \PDB vs.\ \ourTool{} (instrumented) runs.
    \emph{Columns} (left to right): \Exa{}, \Boltz{} solver, \Ewald{} summation.
    \emph{Rows:} \textbf{(a)--(c)}~\Tokyo/NVIDIA RTX~5000 Ada;
    \textbf{(d)--(f)}~\Vista/NVIDIA GH200;
    \textbf{(g)--(i)}~\amdCluster{}/\AMD MI300X.%
  }
  \label{fig:eval-exec-sweeps-combined}
  \vspace{-15pt}

\end{figure*}

\MyPara{\Exa}
Figures~\ref{fig:eval-examinimd-exectime},~\ref{fig:eval-examinimd-exectime-gh200}
and~\ref{fig:eval-examinimd-exectime-amd} show total \Exa execution time for
\PDB and \ourTool debugging sessions across multiple backends as the atom count
increases.
Table~\ref{tab:eval-exec-sweep-summary} reports min--max wall times and
the mean per-point ratio \SlowdownRatio for each configuration;
for \Exa, the mean does not exceed
\UseMacroRound{eval-bench-examinimd-exec-openmp-max-avg-pkdb-slowdown}{2}$\times$
for \OpenMP,
\UseMacroRound{eval-bench-examinimd-exec-cuda-max-avg-pkdb-slowdown}{2}$\times$
for \CUDA, and
\UseMacroRound{eval-bench-examinimd-exec-hip-max-avg-pkdb-slowdown}{2}$\times$
for \HIP{}.

\MyPara{\Boltz}
Figures~\ref{fig:boltzmann-exectime},~\ref{fig:boltzmann-exectime-gh200},
and~\ref{fig:boltzmann-exectime-amd} show that \PDB and \ourTool{} runs
track each other closely on each backend across systems. Table~\ref{tab:eval-exec-sweep-summary}
reports min--max wall times and the mean per-point ratio
\SlowdownRatio for each configuration; for \Boltz, that mean
does not exceed
\UseMacroRound{eval-bench-boltzmann-exec-openmp-max-avg-pkdb-slowdown}{2}$\times$
for \OpenMP,
\UseMacroRound{eval-bench-boltzmann-exec-cuda-max-avg-pkdb-slowdown}{2}$\times$ for
\CUDA, and
\UseMacroRound{eval-bench-boltzmann-exec-hip-max-avg-pkdb-slowdown}{2}$\times$
for \HIP{}.

\MyPara{Ewald}
Figures~\ref{fig:ewald-exectime},~\ref{fig:ewald-exectime-gh200}
and~\ref{fig:ewald-exectime-amd} report total Ewald time for each
configuration. Note that for Figure~\ref{fig:ewald-exectime} we don't have a
runtime for 5M, due to lack of memory on \Tokyo server, which is unrelated to
\ourTool. Table~\ref{tab:eval-exec-sweep-summary} reports min--max wall
times and the mean per-point ratio \SlowdownRatio for each
configuration; for \Ewald, that mean does not exceed
\UseMacroRound{eval-bench-ewald-exec-openmp-max-avg-pkdb-slowdown}{2}$\times$ for
\OpenMP,
\UseMacroRound{eval-bench-ewald-exec-cuda-max-avg-pkdb-slowdown}{2}$\times$
for \CUDA, and
\UseMacroRound{eval-bench-ewald-exec-hip-max-avg-pkdb-slowdown}{2}$\times$
for \HIP{}.

\begin{table}[t]
  \centering
  \caption{%
    Wall-time ranges (seconds) for \Boltz, \Exa, and \Ewald on the \Tokyo,
    \Vista, and \amdCluster node
    (Sec.~\ref{eval:benchmarks-evaluation-section}). Each range is the min--max time
    over all sizes in the corresponding execution-time figures; the last column
    is the mean of $t_i^{\ourTool}/t_i^{\text{pdb}}$ over those same points.}
  \label{tab:eval-exec-sweep-summary}
  \scriptsize
  \setlength{\tabcolsep}{2.5pt}
  \setlength{\extrarowheight}{0.5pt}
  \begin{tabular*}{\columnwidth}{@{}lll@{\extracolsep{\fill}}r@{\extracolsep{\fill}}r@{\extracolsep{\fill}}r@{}}
    \toprule
    \textbf{Benchmark}
      & \textbf{Processor}
      & \textbf{Execution space}
      & \makecell[r]{\textbf{\PDB} \textbf{(s)}}
      & \makecell[r]{\textbf{\ourTool} \textbf{(s)}}
      & \makecell[r]{\textbf{Overhead} ($\times$)} \\
    \midrule
    \multirow{5}{*}{\Exa} & \multirow{2}{*}{Xeon/RTX} & \Openmp & {\UseMacroRound{eval-bench-examinimd-exec-tokyo-openmp-pdb-min}{1}--\UseMacroRound{eval-bench-examinimd-exec-tokyo-openmp-pdb-max}{1}} & {\UseMacroRound{eval-bench-examinimd-exec-tokyo-openmp-pkdb-min}{1}--\UseMacroRound{eval-bench-examinimd-exec-tokyo-openmp-pkdb-max}{1}} & \UseMacroRound{eval-bench-examinimd-exec-tokyo-openmp-avg-pkdb-slowdown}{2} \\
     &  & \Cuda & {\UseMacroRound{eval-bench-examinimd-exec-tokyo-cuda-pdb-min}{1}--\UseMacroRound{eval-bench-examinimd-exec-tokyo-cuda-pdb-max}{1}} & {\UseMacroRound{eval-bench-examinimd-exec-tokyo-cuda-pkdb-min}{1}--\UseMacroRound{eval-bench-examinimd-exec-tokyo-cuda-pkdb-max}{1}} & \UseMacroRound{eval-bench-examinimd-exec-tokyo-cuda-avg-pkdb-slowdown}{2} \\
     & \multirow{2}{*}{GH200} & \Openmp & {\UseMacroRound{eval-bench-examinimd-exec-vista-gh200-openmp-pdb-min}{1}--\UseMacroRound{eval-bench-examinimd-exec-vista-gh200-openmp-pdb-max}{1}} & {\UseMacroRound{eval-bench-examinimd-exec-vista-gh200-openmp-pkdb-min}{1}--\UseMacroRound{eval-bench-examinimd-exec-vista-gh200-openmp-pkdb-max}{1}} & \UseMacroRound{eval-bench-examinimd-exec-vista-gh200-openmp-avg-pkdb-slowdown}{2} \\
     &  & \Cuda & {\UseMacroRound{eval-bench-examinimd-exec-vista-gh200-cuda-pdb-min}{1}--\UseMacroRound{eval-bench-examinimd-exec-vista-gh200-cuda-pdb-max}{1}} & {\UseMacroRound{eval-bench-examinimd-exec-vista-gh200-cuda-pkdb-min}{1}--\UseMacroRound{eval-bench-examinimd-exec-vista-gh200-cuda-pkdb-max}{1}} & \UseMacroRound{eval-bench-examinimd-exec-vista-gh200-cuda-avg-pkdb-slowdown}{2} \\
     & MI300X & \Hip & {\UseMacroRound{eval-bench-examinimd-exec-amd-mi300x-hip-pdb-min}{1}--\UseMacroRound{eval-bench-examinimd-exec-amd-mi300x-hip-pdb-max}{1}} & {\UseMacroRound{eval-bench-examinimd-exec-amd-mi300x-hip-pkdb-min}{1}--\UseMacroRound{eval-bench-examinimd-exec-amd-mi300x-hip-pkdb-max}{1}} & \UseMacroRound{eval-bench-examinimd-exec-amd-mi300x-hip-avg-pkdb-slowdown}{2} \\
    \midrule
    \multirow{5}{*}{\Boltz} & \multirow{2}{*}{Xeon/RTX} & \Openmp & {\UseMacroRound{eval-bench-boltzmann-exec-tokyo-openmp-pdb-min}{1}--\UseMacroRound{eval-bench-boltzmann-exec-tokyo-openmp-pdb-max}{1}} & {\UseMacroRound{eval-bench-boltzmann-exec-tokyo-openmp-pkdb-min}{1}--\UseMacroRound{eval-bench-boltzmann-exec-tokyo-openmp-pkdb-max}{1}} & \UseMacroRound{eval-bench-boltzmann-exec-tokyo-openmp-avg-pkdb-slowdown}{2} \\
     &  & \Cuda & {\UseMacroRound{eval-bench-boltzmann-exec-tokyo-cuda-pdb-min}{1}--\UseMacroRound{eval-bench-boltzmann-exec-tokyo-cuda-pdb-max}{1}} & {\UseMacroRound{eval-bench-boltzmann-exec-tokyo-cuda-pkdb-min}{1}--\UseMacroRound{eval-bench-boltzmann-exec-tokyo-cuda-pkdb-max}{1}} & \UseMacroRound{eval-bench-boltzmann-exec-tokyo-cuda-avg-pkdb-slowdown}{2} \\
     & \multirow{2}{*}{GH200} & \Openmp & {\UseMacroRound{eval-bench-boltzmann-exec-vista-gh200-openmp-pdb-min}{1}--\UseMacroRound{eval-bench-boltzmann-exec-vista-gh200-openmp-pdb-max}{1}} & {\UseMacroRound{eval-bench-boltzmann-exec-vista-gh200-openmp-pkdb-min}{1}--\UseMacroRound{eval-bench-boltzmann-exec-vista-gh200-openmp-pkdb-max}{1}} & \UseMacroRound{eval-bench-boltzmann-exec-vista-gh200-openmp-avg-pkdb-slowdown}{2} \\
     &  & \Cuda & {\UseMacroRound{eval-bench-boltzmann-exec-vista-gh200-cuda-pdb-min}{1}--\UseMacroRound{eval-bench-boltzmann-exec-vista-gh200-cuda-pdb-max}{1}} & {\UseMacroRound{eval-bench-boltzmann-exec-vista-gh200-cuda-pkdb-min}{1}--\UseMacroRound{eval-bench-boltzmann-exec-vista-gh200-cuda-pkdb-max}{1}} & \UseMacroRound{eval-bench-boltzmann-exec-vista-gh200-cuda-avg-pkdb-slowdown}{2} \\
     & MI300X & \Hip & {\UseMacroRound{eval-bench-boltzmann-exec-amd-mi300x-hip-pdb-min}{1}--\UseMacroRound{eval-bench-boltzmann-exec-amd-mi300x-hip-pdb-max}{1}} & {\UseMacroRound{eval-bench-boltzmann-exec-amd-mi300x-hip-pkdb-min}{1}--\UseMacroRound{eval-bench-boltzmann-exec-amd-mi300x-hip-pkdb-max}{1}} & \UseMacroRound{eval-bench-boltzmann-exec-amd-mi300x-hip-avg-pkdb-slowdown}{2} \\
    \midrule
    \multirow{5}{*}{\Ewald} & \multirow{2}{*}{Xeon/RTX} & \Openmp & {\UseMacroRound{eval-bench-ewald-exec-tokyo-openmp-pdb-min}{1}--\UseMacroRound{eval-bench-ewald-exec-tokyo-openmp-pdb-max}{1}} & {\UseMacroRound{eval-bench-ewald-exec-tokyo-openmp-pkdb-min}{1}--\UseMacroRound{eval-bench-ewald-exec-tokyo-openmp-pkdb-max}{1}} & \UseMacroRound{eval-bench-ewald-exec-tokyo-openmp-avg-pkdb-slowdown}{2} \\
     &  & \Cuda & {\UseMacroRound{eval-bench-ewald-exec-tokyo-cuda-pdb-min}{1}--\UseMacroRound{eval-bench-ewald-exec-tokyo-cuda-pdb-max}{1}} & {\UseMacroRound{eval-bench-ewald-exec-tokyo-cuda-pkdb-min}{1}--\UseMacroRound{eval-bench-ewald-exec-tokyo-cuda-pkdb-max}{1}} & \UseMacroRound{eval-bench-ewald-exec-tokyo-cuda-avg-pkdb-slowdown}{2} \\
     & \multirow{2}{*}{GH200} & \Openmp & {\UseMacroRound{eval-bench-ewald-exec-vista-gh200-openmp-pdb-min}{1}--\UseMacroRound{eval-bench-ewald-exec-vista-gh200-openmp-pdb-max}{1}} & {\UseMacroRound{eval-bench-ewald-exec-vista-gh200-openmp-pkdb-min}{1}--\UseMacroRound{eval-bench-ewald-exec-vista-gh200-openmp-pkdb-max}{1}} & \UseMacroRound{eval-bench-ewald-exec-vista-gh200-openmp-avg-pkdb-slowdown}{2} \\
     &  & \Cuda & {\UseMacroRound{eval-bench-ewald-exec-vista-gh200-cuda-pdb-min}{1}--\UseMacroRound{eval-bench-ewald-exec-vista-gh200-cuda-pdb-max}{1}} & {\UseMacroRound{eval-bench-ewald-exec-vista-gh200-cuda-pkdb-min}{1}--\UseMacroRound{eval-bench-ewald-exec-vista-gh200-cuda-pkdb-max}{1}} & \UseMacroRound{eval-bench-ewald-exec-vista-gh200-cuda-avg-pkdb-slowdown}{2} \\
     & MI300X & \Hip & {\UseMacroRound{eval-bench-ewald-exec-amd-mi300x-hip-pdb-min}{1}--\UseMacroRound{eval-bench-ewald-exec-amd-mi300x-hip-pdb-max}{1}} & {\UseMacroRound{eval-bench-ewald-exec-amd-mi300x-hip-pkdb-min}{1}--\UseMacroRound{eval-bench-ewald-exec-amd-mi300x-hip-pkdb-max}{1}} & \UseMacroRound{eval-bench-ewald-exec-amd-mi300x-hip-avg-pkdb-slowdown}{2} \\
    \bottomrule
  \end{tabular*}
\end{table}

\subsection{Comparison with \PyKokkos Debug mode}
\label{eval:debug-time-comparison-section}

\begin{figure}[t]
  \centering
  \footnotesize
  \newcommand{\ExaDebugFigH}{1.2\linewidth}
  \newcommand{\ExaDebugFigW}{1.35\linewidth}

  \ref{fig-examinimd-debug-legend}\par\vspace{0.35ex}
  \begin{minipage}[t]{0.31\columnwidth}
    \centering
    \begin{tikzpicture}
      \begin{axis}[
          width=\ExaDebugFigW,
          height=\ExaDebugFigH,
          xlabel={Atoms},
          ylabel={Time [s]},
          xmin=0.5,
          xmax=5.5,
          xtick={{ 1, 2, 3, 4, 5 }},
          xticklabels={{ 100, 200, 300, 400, 500 }},
          xticklabel style={rotate=45, anchor=east, font=\tiny},
          yticklabel style={font=\tiny},
          xlabel style={font=\small},
          ylabel style={font=\small, at={(axis description cs:-0.22,0.5)},anchor=south},
          legend cell align={left},
          legend to name=fig-examinimd-debug-legend,
          legend style={legend columns=4, font=\scriptsize, draw=none, column sep=0.4em},
          legend entries={\ourTool-OpenMP,\ourTool-Cuda,\ourTool-HIP,\PyKokkos-Debug},
          xmajorgrids=true,
          ymajorgrids=true,
          grid style={dashed,gray!\chartDashedLineOpacity},
          scaled ticks=false,
          ymode=log,
          log basis y={10},
          ymin=-100,
          ymax=3000,
          ytickten={-1,0,1,2,3},
          minor y tick num=0,
          yminorgrids=true,
        ]
        \input{figures/eval-ExaMiniMD-debug-time-data.tex}
      \end{axis}
    \end{tikzpicture}
    \captionof{figure}{\Exa debug wall time (\Tokyo, RTX 5000).}
    \label{fig:examinimd-debug-time-comparison}
  \end{minipage}%
  \hfill%
  \begin{minipage}[t]{0.31\columnwidth}
    \centering
    \begin{tikzpicture}
      \begin{axis}[
          width=\ExaDebugFigW,
          height=\ExaDebugFigH,
          xlabel={Atoms},
          ylabel={},
          xmin=0.5,
          xmax=5.5,
          xtick={{ 1, 2, 3, 4, 5 }},
          xticklabels={{ 100, 200, 300, 400, 500 }},
          xticklabel style={rotate=45, anchor=east, font=\tiny},
          yticklabel style={font=\tiny},
          xlabel style={font=\small},
          xmajorgrids=true,
          ymajorgrids=true,
          grid style={dashed,gray!\chartDashedLineOpacity},
          scaled ticks=false,
          ymode=log,
          log basis y={10},
          ymin=-100,
          ymax=3000,
          ytickten={-1,0,1,2,3},
          minor y tick num=0,
          yminorgrids=true,
        ]
        \input{figures/eval-ExaMiniMD-debug-time-data-gh200.tex}
      \end{axis}
    \end{tikzpicture}
    \captionof{figure}{\Exa debug wall time (\Vista, GH200).}
    \label{fig:examinimd-debug-time-gh200}
  \end{minipage}%
  \hfill%
  \begin{minipage}[t]{0.31\columnwidth}
    \centering
    \begin{tikzpicture}
      \begin{axis}[
          width=\ExaDebugFigW,
          height=\ExaDebugFigH,
          xlabel={Atoms},
          ylabel={},
          xmin=0.5,
          xmax=5.5,
          xtick={{ 1, 2, 3, 4, 5 }},
          xticklabels={{  100, 200, 300, 400, 500  }},
          xticklabel style={rotate=45, anchor=east, font=\tiny},
          yticklabel style={font=\tiny},
          xlabel style={font=\small},
          xmajorgrids=true,
          ymajorgrids=true,
          grid style={dashed,gray!\chartDashedLineOpacity},
          scaled ticks=false,
          ymode=log,
          log basis y={10},
          ymin=-100,
          ymax=3000,
          ytickten={-1,0,1,2,3},
          minor y tick num=0,
          yminorgrids=true,
        ]
        \input{figures/eval-ExaMiniMD-debug-time-data-amd.tex}
      \end{axis}
    \end{tikzpicture}
    \captionof{figure}{\Exa debug wall time (\amdCluster, \Hip).}
    \label{fig:examinimd-debug-time-amd}
  \end{minipage}

\end{figure}

\noindent
Figures~\ref{fig:examinimd-debug-time-comparison},~\ref{fig:examinimd-debug-time-gh200},
and~\ref{fig:examinimd-debug-time-amd} compare \Exa{} debug wall time for two
strategies. \PyKokkos-Debug denotes the existing debug execution mode: parallel
kernels are lowered to sequential, Python-only execution and debugged with
\PDB---a common \Python \eDSL debugging
approach~\cite{tritonDebugging,numbaCudaSim}---and compared with \ourTool{}
debugging with \CUDA, \OpenMP, and \HIP{} backends, where we observe significant
speedups over \PyKokkos-Debug.

We plot timings across backends for atom counts from 100 up to 500; beyond that
range the \PyKokkos-Debug quickly becomes impractical because each
timestep runs sequentially on the host, inside the \Python interpreter.
Across the full size range, \PyKokkos-Debug spans
\UseMacroRound{eval-debugtime-tokyo-baseline-t-min}{2}--\UseMacroRound{eval-debugtime-tokyo-baseline-t-max}{2}s
on \Tokyo and
\UseMacroRound{eval-debugtime-vista-gh200-baseline-t-min}{2}--\UseMacroRound{eval-debugtime-vista-gh200-baseline-t-max}{2}s
on \Vista, while \ourTool-\OpenMP stays between
\UseMacroRound{eval-debugtime-tokyo-pkdbopenmp-t-min}{2} and
\UseMacroRound{eval-debugtime-tokyo-pkdbopenmp-t-max}{2}s on \Tokyo (%
\UseMacroRound{eval-debugtime-vista-gh200-pkdbopenmp-t-min}{2}--\UseMacroRound{eval-debugtime-vista-gh200-pkdbopenmp-t-max}{2}s
on \Vista), and \ourTool-\CUDA between
\UseMacroRound{eval-debugtime-tokyo-pkdbcuda-t-min}{2} and
\UseMacroRound{eval-debugtime-tokyo-pkdbcuda-t-max}{2}s on \Tokyo (%
\UseMacroRound{eval-debugtime-vista-gh200-pkdbcuda-t-min}{2}--\UseMacroRound{eval-debugtime-vista-gh200-pkdbcuda-t-max}{2}s
on \Vista); on \amdCluster, \PyKokkos-Debug spans
\UseMacroRound{eval-debugtime-amd-mi300x-baseline-t-min}{2}--\UseMacroRound{eval-debugtime-amd-mi300x-baseline-t-max}{2}s
and \ourTool-\HIP stays between
\UseMacroRound{eval-debugtime-amd-mi300x-pkdbhip-t-min}{2} and
\UseMacroRound{eval-debugtime-amd-mi300x-pkdbhip-t-max}{2}s.

\subsection{\KernelSubstitute cost}
\label{eval:hot-swapping-section}

To evaluate \kernelSubstitute, we modified the \Exa benchmark by adding four
alternate workunits that a programmer could swap in at the \parallelOpsCode
call site\footnote{We choose one call site as \textbf{\EditDebug} scales
      linearly while \textbf{call site substitution} has constant asymptotic time
      (given infinite processes).}.
The alternate kernels are:

\begin{enumerate}[leftmargin=*, itemsep=1pt, topsep=3pt, parsep=0pt, partopsep=0pt]
      \item \textbf{Temperature reduction}: \CodeIn{compute\_workunit} swapped
            for \CodeIn{compute\_workunit\_half}, which is a half-scaled kernel.
      \item \textbf{Energy reduction}: \CodeIn{work} swapped for
            \CodeIn{work\_half}, which is a half-scaled kernel.
      \item \textbf{First integrator}: \CodeIn{initial\_integrate} swapped
            for \linebreak~\CodeIn{initial\_integrate\_clean} (zeros \CodeIn{v}
            and \CodeIn{x} for probing).
      \item \textbf{Last integrator}: \CodeIn{final\_integrate} swapped
            for \linebreak~\CodeIn{final\_integrate\_clean} (zeros \CodeIn{v} for probing).
\end{enumerate}

\noindent
Regarding size, we picked median value from our sample, which is 420k. All
kernels were tested using \Cuda execution space and compared against \PDB
\EditDebug sessions.

\begin{table}[t]
    \centering
    \setlength{\tabcolsep}{4pt}
    \caption{ExaMiniMD \kernelSubstitute benchmark with atom size of 420k,
    executed on \Tokyo/\RTXFiveKAda. All times in seconds except
    $t_{\mathrm{hs}}$ (milliseconds): $t_{A}$ is two-run
    edit-and-debug with \PDB total; $t_{A1}$, $t_{A2}$ are the first and second
    run of \PDB sessions; $t_{B}$ uses~\CodeIn{hotswap} during \ourTool
    interactive session; $t_{\mathrm{hs}}$ is substitution command latency.}
    \label{tab:eval-hotswap-benchmark}
    \begin{tabular}{lrrrrrr}
        \toprule
        \textbf{Preset} & \textbf{$t_A$ (s)} & \textbf{$t_{A1}$ (s)} & \textbf{$t_{A2}$ (s)} & \textbf{$t_B$ (s)} & \textbf{Spd. ($\times$)} & \textbf{$t_{\mathrm{hs}}$ (ms)} \\
\midrule
\UseMacro{eval-hotswap-temperature-name} &
\UseMacro{eval-hotswap-temperature-tA} &
\UseMacro{eval-hotswap-temperature-tA1} &
\UseMacro{eval-hotswap-temperature-tA2} &
\UseMacro{eval-hotswap-temperature-tB} &
\UseMacro{eval-hotswap-temperature-speedup} &
\UseMacro{eval-hotswap-temperature-hotswap-cmd} \\
\UseMacro{eval-hotswap-kine-name} &
\UseMacro{eval-hotswap-kine-tA} &
\UseMacro{eval-hotswap-kine-tA1} &
\UseMacro{eval-hotswap-kine-tA2} &
\UseMacro{eval-hotswap-kine-tB} &
\UseMacro{eval-hotswap-kine-speedup} &
\UseMacro{eval-hotswap-kine-hotswap-cmd} \\
\UseMacro{eval-hotswap-nve-initial-name} &
\UseMacro{eval-hotswap-nve-initial-tA} &
\UseMacro{eval-hotswap-nve-initial-tA1} &
\UseMacro{eval-hotswap-nve-initial-tA2} &
\UseMacro{eval-hotswap-nve-initial-tB} &
\UseMacro{eval-hotswap-nve-initial-speedup} &
\UseMacro{eval-hotswap-nve-initial-hotswap-cmd} \\
\UseMacro{eval-hotswap-nve-final-name} &
\UseMacro{eval-hotswap-nve-final-tA} &
\UseMacro{eval-hotswap-nve-final-tA1} &
\UseMacro{eval-hotswap-nve-final-tA2} &
\UseMacro{eval-hotswap-nve-final-tB} &
\UseMacro{eval-hotswap-nve-final-speedup} &
\UseMacro{eval-hotswap-nve-final-hotswap-cmd} \\
\midrule
\textbf{\UseMacro{eval-hotswap-total-name}} &
\UseMacro{eval-hotswap-total-tA} &
\UseMacro{eval-hotswap-total-tA1} &
\UseMacro{eval-hotswap-total-tA2} &
\UseMacro{eval-hotswap-total-tB} &
\UseMacro{eval-hotswap-total-speedup} &
\UseMacro{eval-hotswap-total-hotswap-cmd} \\
\bottomrule
    \end{tabular}
\end{table}

\noindent
We compare the efficiency of our call site substitution implementation
to relaunching separate debugging sessions after each on-disk edit,
using a fully automated experiment:
\begin{enumerate*}
      \item launch \ourTool in a subprocess;
      \item set breakpoints before and after the kernel call;
      \item run \textbf{\EditDebug}---save changes on disk and launch two separate \ourTool sessions;
      \item run \textbf{call site substitution}---launch a single session that issues \CodeIn{hotswap} at the call site;
      \item record wall-clock intervals and verify debugger outputs.
\end{enumerate*}

Results are reported in Table~\ref{tab:eval-hotswap-benchmark}.
We target kernels that run early in each \Exa timestep so both cases
reach the inspection point after modest work\footnote{Setting a breakpoint later in the execution would increase \textbf{\EditDebug}'s runtime but have no effect on \textbf{call site substitution} runtime.}.
Both cases exit immediately after the kernel returns.

As expected, \textbf{\EditDebug} is more expensive than \textbf{call site
      substitution}, as it launches a second \ourTool session and reruns the program
up to the call sites after each edit. The \CodeIn{hotswap} command itself adds
only tens of milliseconds of wall time (Table~\ref{tab:eval-hotswap-benchmark},
$t_{\mathrm{hs}}$), a negligible overhead.

\subsection{Case studies}
\label{eval:case-studies-section}

We use \ourTool to debug the \Boltz and \Ewald \PyKokkos applications.

\subsubsection{\Boltz}

\begin{figure}[t]
  \centering
  \lstset{
    basicstyle=\ttfamily\scriptsize,
    breaklines=true,
    breakatwhitespace=false,
    showstringspaces=false,
    frame=single,
    framerule=0.4pt,
    xleftmargin=2em,
    xrightmargin=0.6em,
    aboveskip=4pt,
    belowskip=4pt,
    columns=fullflexible,
  }
  \begin{lstlisting}[language={},escapechar=|]
cudaDeviceSynchronize() error( cudaErrorIllegalAddress): an illegal memory access .../Cuda/Kokkos_Cuda_Instance.cpp:154 |\label{boltzmann:illegalAddress}|
Backtrace:
[0x7d...c9] Kokkos::Impl::save_stacktrace()
[0x7d...50] Kokkos::Impl::host_abort(char const*)
[0x7d...bb] Kokkos::Impl::cuda_internal_error_abort(...)
[0x7d...1a] Kokkos::Impl::cuda_device_synchronize(...)
[0x7d...0d] Kokkos::Impl::ExecSpaceManager::static_fence(...)
[0x7d...d1] void Kokkos::deep_copy<...>(...)
[0x7d...0e] run(...)
|\hfill\makebox[0pt][c]{\ttfamily\scriptsize\ldots}\hfill\null|
[0x5a...3e] _start
Aborted (core dumped)
\end{lstlisting}
  \caption{Boltzmann benchmark: \texttt{cudaErrorIllegalAddress} at \texttt{cudaDeviceSynchronize} (Kokkos \texttt{deep\_copy} between \texttt{CudaSpace} and \texttt{CudaUVMSpace}).}
  \label{fig:boltzmann-cuda-error}
\end{figure}

While developing and evaluating \ourTool on the \Boltz benchmark, we encountered
a bug, which we then debugged using \ourTool. The \Boltz benchmark failed during
\CUDA execution: the run aborted with \CodeIn{cudaErrorIllegalAddress} on the
line~\ref{boltzmann:illegalAddress} in Figure~\ref{fig:boltzmann-cuda-error}.
Here, the argument memory spaces and layouts looked consistent with a legal
launch, yet we could only stop on the first line of the workunit and step
through to line~\ref{boltzmann:bb-error}
(Figure~\ref{fig:boltzmann-redcheck-unfixed}), after which the benchmark crashed
with the error messages in Figure~\ref{fig:boltzmann-warp-illegal-address}.

\begin{figure}[t]
  \centering
  \lstset{
    language=Python,
    basicstyle=\ttfamily\scriptsize,
    keywordstyle=\bfseries,
    numbers=left,
    numbersep=6pt,
    showstringspaces=false,
    breaklines=true,
    frame=single,
    framerule=0.4pt,
    xleftmargin=1.5em,
    xrightmargin=0.6em,
    aboveskip=4pt,
    belowskip=4pt,
    literate={@}{{\char`@}}1 {->}{{$\rightarrow$}}2,
  }
  \begin{lstlisting}[language={python},escapechar=|]
@pk.workunit(scratch=[(float, lambda p, s: s.scratch_s)])|\label{boltzmann:workunit-definition}|
def redcheck(self, t: pk.TeamMember): |\label{boltzmann:function-definition}|
    i: int = t.league_rank() * t.team_size() + t.team_rank()
    tid: int = t.team_rank()
    if i < self.M:
        bb: pk.ScratchView1D[float] = pk.ScratchView1D(t.team_scratch(0))
        for j in range(self.r):
            bb[self.r*tid + j] = self.w[self.r*i + j]|\label{boltzmann:bb-error}|
        t.team_barrier()
  \end{lstlisting}
  \caption{Boltzmann \CodeIn{redcheck} workunit definition under debugging.}
  \label{fig:boltzmann-redcheck-unfixed}
\end{figure}

\begin{figure}[t]
  \centering
  \lstset{
    basicstyle=\ttfamily\scriptsize,
    breaklines=true,
    breakatwhitespace=false,
    showstringspaces=false,
    frame=single,
    framerule=0.4pt,
    xleftmargin=1.5em,
    xrightmargin=0.6em,
    aboveskip=4pt,
    belowskip=4pt,
    columns=fullflexible,
  }
  \begin{lstlisting}[language={}]
[pkdb-cuda] accelerator debugger: CUDA Exception: Warp Illegal Address
[pkdb-cuda] accelerator debugger: The exception was triggered at PC 0x7c..10  pk_functor_Scale<Kokkos::Cuda>::operator()(...) const  (functor.hpp:37)
[pkdb-cuda] accelerator debugger: Thread 15 "cuda-EvtHandlr" received signal CUDA_EXCEPTION_14, Warp Illegal Address.
  \end{lstlisting}
  \caption{Boltzmann session: pkdb \CUDA exception reporting warp illegal address at generated functor code (\texttt{functor.hpp:37}).}
  \label{fig:boltzmann-warp-illegal-address}
\end{figure}

This error message suggested that the scratch size was not correctly specified;
we discovered that the \PyKokkos scratch size specification via the lambda definition on line~\ref{boltzmann:workunit-definition} was broken:
\PyKokkos correctly registered the scratch size, but failed to translate the
scratch specification to C++. As \CodeIn{pk.TeamMember} on
line~\ref{boltzmann:function-definition} was translated to the C++, we used
native debugger-oriented command \CodeIn{send print this->team\_size}, which
shows size of current team of threads, and observed that the team size was set
to $0$. Thus, the cache was not allocated for the team, since there is no team
at all. After resolving this issue in \PyKokkos, the \Boltz benchmark ran
successfully.

\subsubsection{\Ewald}

\def\pktranscriptHLY#1{\colorbox{yellow!35}{\strut\ttfamily\scriptsize\detokenize{#1}}}%
\def\pktranscriptHLG#1{\colorbox{green!22}{\strut\ttfamily\scriptsize\detokenize{#1}}}%

\begin{figure}[t]
      \centering
      \lstset{
            basicstyle=\ttfamily\scriptsize,
            breaklines=true,
            breakatwhitespace=false,
            showstringspaces=false,
            frame=single,
            framerule=0.4pt,
            xleftmargin=2em,
            xrightmargin=0.6em,
            aboveskip=4pt,
            belowskip=4pt,
            columns=fullflexible,
      }

      \begin{subfigure}{\linewidth}
            \centering
            \begin{lstlisting}[language={},escapechar=|]
Segmentation fault         (core dumped) python celllist.py --device OpenMP|\label{parki:openmp-segfault}|
    \end{lstlisting}
            \caption{Host OpenMP: \CodeIn{celllist.py} aborts with no Python traceback.}
            \label{fig:parki-openmp-segfault}
      \end{subfigure}

      \vspace{0.6em}

      \begin{subfigure}{\linewidth}
            \centering
            \begin{lstlisting}[language={},escapechar=|]
cudaStreamSynchronize(stream) error( cudaErrorIllegalAddress): an illegal memory access ../Cuda/Kokkos_Cuda_Instance.cpp:165|\label{parki:cuda-illegalAddress}|
Backtrace:
[0x72..19] Kokkos::Impl::save_stacktrace()
[0x72..d0] Kokkos::Impl::host_abort(char const*)
|\hfill\makebox[0pt][c]{\ttfamily\scriptsize\ldots}\hfill\null|
[0x72..39] reshuffle_particles_fp64(*kwargs)      |\label{parki:reshuffle-frame}|
|\hfill\makebox[0pt][c]{\ttfamily\scriptsize\ldots}\hfill\null|
[0x72..40] __libc_start_main
[0x5d..3e] _start
Aborted                    (core dumped) python celllist.py --device Cuda
    \end{lstlisting}
            \caption{CUDA: Kokkos abort at \CodeIn{cudaStreamSynchronize} after \CodeIn{reshuffle\_particles\_fp64}.}
            \label{fig:parki-cuda-illegal}
      \end{subfigure}

      \vspace{0.6em}

      \begin{subfigure}{\linewidth}
            \centering
            \lstset{
            language=Python,
            basicstyle=\ttfamily\scriptsize,
            keywordstyle=\bfseries,
            numbers=left,
            numbersep=5pt,
            showstringspaces=false,
            breaklines=true,
            frame=single,
            framerule=0.4pt,
            xleftmargin=2em,
            xrightmargin=0.6em,
            aboveskip=4pt,
            belowskip=4pt,
            literate={@}{{\char`@}}1 {->}{{$\rightarrow$}}2,
            }
            \begin{lstlisting}[language=Python,escapechar=|]
@pk.function
def _get_cell_fp64(p, ...) -> int:
    ...
    cell_xyz = [p[0][i] / c_size[0], ..., ...] |\label{line:parki-getcell-xyz}|
    return get_idx(cell_xyz)

@pk.workunit
def reshuffle_particles_fp64(i, counter, p, ...):
    |\pktranscriptHLG{cell: int = _get_cell_fp64(p, ...)}|    |\label{line:parki-reshuffle-get-cell-fp64}|
    |\pktranscriptHLY{offset: int = pk.atomic_add(counter, [cell], 1)}||\label{line:parki-reshuffle-atomic-fetch}|
    ...
    \end{lstlisting}
            \caption{Illustrative part of (\CodeIn{celllist\_workunits.py}): cell indexing and reshuffle path.}
            \label{fig:parki-celllist-code}
      \end{subfigure}

      \vspace{0.6em}

      \begin{subfigure}{\linewidth}
            \centering
            {\setlength{\fboxsep}{0.2em}%
                  \begin{lstlisting}[language={},escapechar=!]
(pkdb) print _grid_shape
!\pktranscriptHLG{array([9, 8, 7])}!
(pkdb) continue
Workunit breakpoint, celllist_workunits.py:!\ref{line:parki-reshuffle-get-cell-fp64}!
!\pktranscriptHLY{(pkdb-device) print len(counter)}!   !\label{parki-print-len}!
!\pktranscriptHLY{int(504)}!             !\label{parki-incorrect-index-504}!
(pkdb-device) continue
Workunit breakpoint, celllist_workunits.py:!\ref{line:parki-getcell-xyz}!
(pkdb-device) parallel_print cell_xyz
tid | cell_xyz
--------------
!\pktranscriptHLG{0   | [9, 7, 0]}\label{parki-incorrect-index-9}!
!\dots!
!\pktranscriptHLG{9   | [2, 8, 2]}\label{parki-incorrect-index-8}!
(pkdb-device) continue
Workunit breakpoint, celllist_workunits.py:!\ref{line:parki-reshuffle-atomic-fetch}!
(pkdb-device) print cell                 !\label{parki-print-cell}!
!\pktranscriptHLY{int(528)}!             !\label{parki-incorrect-cell-528}!


    \end{lstlisting}
            }
            \caption{pkdb session: host-sided \CodeIn{print \_grid\_shape}, then workunit stops at
                  \CodeIn{celllist\_workunits.py}
                  lines~\ref{line:parki-getcell-xyz} and~\ref{line:parki-reshuffle-get-cell-fp64}, and \CodeIn{parallel\_print} of
                  \CodeIn{cell\_xyz} per thread.}
            \label{fig:parki-debug-transcript}
      \end{subfigure}

      \caption{Ewald case study: failure logs (a--b); broken
            \CodeIn{reshuffle}/\CodeIn{\_get\_cell} code with lines of interest
            marked (c); and a \ourTool transcript stopping at those lines (d).}
      \label{fig:parki-case-errors}
\end{figure}

During \Ewald benchmark execution for \ourTool timing comparison, we encountered
a non-deterministic failure with a cryptic error message for \CUDA builds and
silent fails for \OpenMP builds. Figure~\ref{fig:parki-case-errors} shows
console logs (\ref{fig:parki-openmp-segfault}, \ref{fig:parki-cuda-illegal}),
the relevant \texttt{celllist\_workunits.py} listing with marked
lines~(\ref{fig:parki-celllist-code}), and a \ourTool
transcript~(\ref{fig:parki-debug-transcript}). The \OpenMP run
(line~\ref{parki:openmp-segfault} in Figure~\ref{fig:parki-openmp-segfault})
terminates with only a shell-level segmentation fault, whereas the \CUDA run
(Figure~\ref{fig:parki-cuda-illegal}) surfaces \CodeIn{cudaErrorIllegalAddress}
at \CodeIn{cudaStreamSynchronize} with \CodeIn{reshuffle\_particles\_fp64}
on the stack (l
ine~\ref{parki:reshuffle-frame}).

Guided by the \CUDA backtrace in Figure~\ref{fig:parki-cuda-illegal}, we set a
breakpoint inside of \CodeIn{reshuffle\_particles\_fp64} workunit  on the
line~\ref{line:parki-reshuffle-get-cell-fp64} in
Figure~\ref{fig:parki-celllist-code}, and stepped through the launch path.
Evaluating Python-side expressions in \ourTool{} (e.g., applying
\CodeIn{len(<arg>)} to the arguments) revealed out-of-bounds indexing.
Lines~\ref{parki-print-len}, \ref{parki-incorrect-index-504} in
Figure~\ref{fig:parki-debug-transcript} show the length of array, while
lines~\ref{parki-print-cell},~\ref{parki-incorrect-cell-528} show that \Ewald is
trying access element with~\EwaldIndex index.

Continuing, we discovered a bug in ~\CodeIn{\_get\_cell\_fp64} (Figure~\ref{fig:parki-celllist-code}, line~\ref{line:parki-getcell-xyz}), which computes
the cell index; the function
mishandles the case where a particle lies exactly on a cell boundary.
Figure~\ref{fig:parki-debug-transcript} shows a representative \ourTool trace: after
printing \CodeIn{\_grid\_shape}, we stop first at
line~\ref{line:parki-reshuffle-get-cell-fp64},
Figure~\ref{fig:parki-celllist-code} (the \CodeIn{\_get\_cell} call in
\CodeIn{reshuffle\_particles\_fp64}) and then at
line~\ref{line:parki-getcell-xyz}, Figure~\ref{fig:parki-celllist-code} inside
the helper.
We issue the \CodeIn{parallel\_print}
instruction at the device breakpoint to print \CodeIn{cell\_xyz}. The
\CodeIn{parallel\_print} evaluates
the expression once per accelerator thread and prints the resulting values as a
thread-indexed list.
We then continue to the following line in the reshuffle workunit
(Figure~\ref{fig:parki-celllist-code}, line~\ref{line:parki-reshuffle-atomic-fetch}),
where \CodeIn{pk.atomic\_add} updates \CodeIn{counter} at the computed cell
index, and observe that this index is out of bounds for \CodeIn{counter} (yellow
highlight in the figure).

\ourTool was able to help us find a bug in the \Ewald research code, not a
synthetic regression harness; leading the developers to the exact location of
the bug to be fixed.

This case study is a natural fit for \HPC-oriented debugging: ~\ourTool{} can
issue device-side instructions (e.g., \CodeIn{parallel\_print <args>}), and
inspect live per-thread state in one session, rather than being limited to a
single active host thread when a parallel kernel misbehaves.

\section{Limitations and Future Work}

Like most full-featured debuggers, \ourTool trades raw performance for
observability: end-to-end runs with the debugger-instrumented \Kokkos build are
slower than with the paired release build. However, the performance overhead,
based on our evaluation (\ref{eval:benchmarks-evaluation-section}), is moderate.

We selected \PyKokkos as a proof-of-concept framework.
However, we believe that the core ideas behind PKDB generalize to other eDSLs
(such as Triton~\cite{TritonLang}, \Numba~\cite{numbaPaper},
Pallas~\cite{pallasPaper}) and other mixed-language stacks, e.g., \Python front
ends that invoke native \CUDA, \HIP, or \OpenMP code through \pybind or similar
bindings, but substantial engineering might be required.

Our future work involves several research directions. The first is user
experience, which requires conducting user study against existing debugging
methods, such as print-debugging and \PDB-based debugging with no access to
kernel code. Another direction is focused on developing an \eDSL-agnostic layer
of~\ourTool, which would provide the flexible support without \eDSL
dependencies.

\section{Related Work}
\label{sec:related}

We survey prior work relevant to~\ourTool:
\begin{enumerate*}
      \item interactive debugging for \HPC and \GPU programs;
      \item debugging for Python \eDSL and multi-language stacks;
      \item dynamic software updating and \kernelSubstitute; and
      \item performance-portable programming frameworks.
\end{enumerate*}

\MyPara{\HPC and \GPU debugging} Interactive debugging at scale has a long
history in the \HPC community \cite{francioniHPCDebuggingStandard}. Prior
works~\cite{debugHPC,debugMillionCores,WISMULLER1996415} study the challenges of
debugging massively parallel applications without disrupting program execution.
\cite{Knobloch_Mohr_2020} surveys debugging tools for \GPU applications,
including the native debuggers \CUDAGDB~\cite{cudagdbDocumentation} and
\ROCGDB~\cite{ROCgdbDocumentation} that \ourTool uses internally;
\cite{Kokkos_profile} provides \Kokkos hooks for the aforementioned device
debuggers. All of these tools target native C++ or \CUDA code; none provide an
interactive debugging interface for \Python \eDSL{} kernels. \ourTool fills this
gap by bridging \PDB \Python debugging with preexisting device debuggers with
uniform interface to the developer, and bringing novel debugging features (e.g.,
\liveeval, \kernelSubstitute).

\MyPara{\Python \eDSL and multi-language debugging}
A common debugging approach for \Python \GPU \eDSL{}s is to fall back
to \CPU execution.
\Triton~\cite{TritonLang, tritonDebugging} and \Numba~\cite{numbaCudaDebug,
      numbaCudaSim} provide \CPU simulators to step through kernels on
the \CPU; masking device-specific bugs \cite{NumericalBugsInGPU, zhanallclose}
and requiring alteration of the program or its data to keep execution time
manageable. Alternatively, \PyCuda~\cite{pyCudaDebugging} allows users to step
through generated \CUDA kernels using default~\CUDAGDB, but all \Python context,
LOC mapping, and variable translation must be handled manually by the user,

Multi-language debugging is a closely related challenge \cite{9246706, 10.1145/3540250.3549173, 10413900}.
\cite{BlinkDebugger}
compose a debugger across Java and C; similar tools exist for Cython
\cite{florisson_multilevel}, a compiled language providing C-extensions for
\Python. \cite{pavletic2016interactive} develops a framework for debugging more
general extensible C languages, like~\mbeddr~\cite{Mbeddr}. \ourTool follows a similar
composition strategy: \pdbp interposes on \PyKokkos dispatch, handing control to
a platform-specific controller that manages the target debugger. Unlike
\cite{BlinkDebugger,florisson_multilevel,pavletic2016interactive}, \ourTool
targets performance-portable kernels, supports concurrent execution on
accelerators and introduces capabilities such as \liveeval and
\kernelSubstitute.

\MyPara{Dynamic software updating and \kernelSubstitute}
Dynamic software updating (DSU) \cite{onFlightPatching,287718} \cite{5386829}
replaces running code without stopping a live process.
These concepts are crucial to modern debugging; e.g., the Wolverine
debugger~\cite{wolverineDebugRepair} uses these tools to diagnose and repair C
programs without restarting. Recent work extends DSU to active long-running
functions~\cite{stromback2024activedsu} and whole-kernel
subsystems~\cite{ma2023plugsched}. \ourTool's \kernelSubstitute is motivated by the same
desire to avoid costly restarts, but targets \PyKokkos kernels specifically:
replacements are recorded as call-site substitutions that take effect at the
next \parallelOpsCode dispatch, and the substitution semantics are tied to
original call sites rather than to inter-component
dependencies~\cite{hayden2014kitsune, stromback2024activedsu, mvedsua2019asplos}
(Section~\ref{tech:kernel-hotswapping-section}).
\MyPara{Performance-portable and Python HPC frameworks}
\ourTool targets \PyKokkos~\cite{PyKokkos}, which translates a Python
\eDSL to \Kokkos C++~\cite{kokkos,KokkosEcosystem2021} and dispatches
kernels to device backends.  Related performance-portable frameworks
include RAJA~\cite{8945721}, Leonid~\cite{10.1145/3721145.3728489} and ORCHA~\cite{ORCHA, ORCHA2}.
\Numba~\cite{numbaPaper} and JAX~\cite{jax2018github} represent
alternative approaches to Python \HPC, JIT-compiling kernels to LLVM or
XLA; both lack interactive on-device debugging.
Legion \cite{legion_2012} and its \Python \eDSL Pygion \cite{pygion_2019}
provide a performance-portable task-based parallelization framework for \emph{distributed} computations.
The Scout DSL \cite{legion_debugging_2014} is built on top of~\Legion and supports a custom debugger based off LLDB, but as far as we are aware this debugger has \emph{not} been extended to support Legion or Pygion itself.
\ourTool's architecture is designed to be extensible: adding
support for a new backend---whether another \PyKokkos execution space or
a different Python \eDSL---requires implementing the controller
interface described in \S~\ref{tech:arch-overview-section}.

\section{Conclusions}

We presented \ourTool, the first interactive debugger for
performance-portable \Python \HPC kernels.  \ourTool brings familiar
\CodeIn{pdb}-style de\-bugging---breakpoints, stepping, and variable
inspection---to kernels executing natively on \CPUs and \GPUs, without
source modification.  Beyond standard debugging, \ourTool introduces
\liveeval, which lets developers execute arbitrary code directly on a
live device during a paused session, and \kernelSubstitute, which allows
kernels to be updated and reloaded without restarting the application.
We hope \ourTool lowers the barrier to productive development of
performance-portable \HPC software in \Python, and serves as a
foundation for richer tooling in this space, which traditionally has not received sufficient attention by researchers.

\section*{Acknowledgment}

We thank Jakob Bludau, Damien Lebrun-Grandie, and the anonymous reviewers for
their advice and helpful feedback. This work was partially funded by the U.S.
Department of Energy, National Nuclear Security Administration Award Number
DE-NA0003969; the U.S. National Science Foundation (NSF) Nos. CCF-2217696,
CCF-2313027, CCF-2403036; and AMD (University Program AI \& HPC Cluster). Any
opinions, findings, and conclusions or recommendations expressed in this
material are those of the authors and do not necessarily reflect the views of
the funding entities.

\bibliographystyle{IEEEtran}
\bibliography{bib}

\end{document}